\documentclass[preprint]{aastex61}

\usepackage{graphicx}

\newcommand{\asec}{\hbox to 1pt{}\rlap{$^{\prime\prime}$}.\hbox to 2pt{}}
\newcommand{\amin}{\hbox to 1pt{}\rlap{$^{\prime}$}.\hbox to 1pt{}}
\newcommand{\adeg}{\hbox to 1pt{}\rlap{$^{\circ}$}.\hbox to 2pt{}}

\submitjournal{{\it Icarus}}
\shortauthors{Lauer et al.}
\shorttitle{New Horizons and HST Pluto Ring Search}

\begin{document}

\title{The New Horizons and Hubble Space Telescope Search For
Rings, Dust, and Debris in the Pluto-Charon System}

\author{Tod R. Lauer}
\affil{National Optical Astronomy Observatory,\footnote{The
National Optical Astronomy Observatory is operated by AURA, Inc
under cooperative agreement with NSF.}
P.O. Box 26732, Tucson, AZ 85726}

\author{Henry B. Throop}
\affil{Planetary Science Institute, 1700 E Fort Lowell Rd. \#106,
Tucson, AZ 85719}

\author{Mark R. Showalter}
\affil{SETI Institute, Mountain View, CA 94043}

\author{Harold A. Weaver}
\affil{The Johns Hopkins University Applied Physics Laboratory,
Laurel, MD 20723-6099}

\author{S. Alan Stern}
\affil{Department of Space Studies, Southwest Research Institute,
1050 Walnut St., Suite 300, Boulder, CO 80302}

\author{John R. Spencer}
\affil{Department of Space Studies, Southwest Research Institute,
1050 Walnut St., Suite 300, Boulder, CO 80302}

\author{Marc W. Buie}
\affil{Department of Space Studies, Southwest Research Institute,
1050 Walnut St., Suite 300, Boulder, CO 80302}

\author{Douglas P. Hamilton}
\affil{Astronomy Department, University of Maryland, College Park, MD 20742}

\author{Simon B. Porter}
\affil{Department of Space Studies, Southwest Research Institute,
1050 Walnut St., Suite 300, Boulder, CO 80302}

\author{Anne J. Verbiscer}
\affil{Department of Astronomy, University of Virginia, Charlottesville, VA
22904}

\author{Leslie A. Young}
\affil{Department of Space Studies, Southwest Research Institute,
1050 Walnut St., Suite 300, Boulder, CO 80302}

\author{Cathy B. Olkin}
\affil{Department of Space Studies, Southwest Research Institute,
1050 Walnut St., Suite 300, Boulder, CO 80302}

\author{Kimberly Ennico}
\affil{NASA Ames Research Center, Moffett Field, CA 94035}

\author{the New Horizons Science Team}

\begin{abstract}
We conducted an extensive search for dust or debris rings in the Pluto-Charon system
before, during, and after the New Horizons encounter in July 2015.
Methodologies included attempting to detect features by
back-scattered light during the approach to Pluto (phase angle $\alpha\sim15^\circ$),
{\it in situ} detection of impacting particles, a search for
stellar occultations near the time of closest approach,
and by forward-scattered light imaging during departure ($\alpha\sim165^\circ$).
An extensive search using the Hubble Space Telescope (HST) prior
to the encounter also contributed to the final ring limits.
No rings, debris, or dust features were observed,
but our new detection limits provide a substantially
improved picture of the environment throughout the Pluto-Charon system.
Searches for rings in back-scattered light were covered the 
range 35,000--250,000 km from the system barycenter,  a zone that starts 
interior to the orbit of Styx, the innermost minor satellite, and extends
out to four times the orbital radius of Hydra, the outermost known satellite.
We obtained our firmest limits using
data from the New Horizons LORRI camera in the inner half of this region.
Our limits on the normal $I/F$ of an unseen ring depends on the radial scale of the rings:
$2\times10^{-8}$ ($3\sigma$) for 1500 km wide rings, $1\times10^{-8}$ for 6000 km rings, and 
$7\times10^{-9}$ for 12,000 km rings.
Beyond $\sim100,000$~km from Pluto, HST observations limit normal $I/F$ to
$\sim8\times10^{-8}$.
Searches for dust features from forward-scattered light extended from the
surface of Pluto to the Pluto-Charon Hill sphere ($r_{\rm Hill}=6.4\times10^6$ km).
No evidence for rings or dust clouds was detected to normal
$I/F$ limits of $\sim8.9\times10^{-7}$ on $\sim10^4$ km scales.
Four stellar occulation observations also probed the
space interior to Hydra, but again no dust or debris
was detected. The Student Dust Counter detected one particle impact
$3.6\times10^6$ km from Pluto, but this is consistent
with the interplanetary space environment established
during the cruise of New Horizons.
Elsewhere in the solar system, small moons commonly share their orbits with
faint dust rings.
Our results support recent dynamical studies suggesting that small grains
are quickly lost from the Pluto-Charon system due to solar radiation pressure,
whereas larger particles are orbitally unstable due to ongoing
perturbations by the known moons.

\end{abstract}


\section{Searching For Rings and Dust Around Pluto}

\subsection{Could Pluto Have Rings?}

It is likely that Pluto had rings \footnote{For brevity,
we generally describe the search for all such features
as a search for ``rings'' throughout the paper, even though the
observations and analysis are sensitive to diffuse or extended
debris or dust features as well.} at various times in its history,
although their existence may have been fleeting.
The standard model for the formation of
both Charon and the four minor satellites of Pluto is that they were
created in a collision between Pluto and another large KBO
\citep{charon, canup, stern06},
which would have created an extensive debris disk.
Dynamical interactions would have quickly cleared most of the larger debris,
and solar radiation pressure sweeps away much of the fine dust \citep{solar};
however, it is plausible that remnants of this initial event
could have persisted at large radii into the present \citep{kenyon}.
The recent discovery of rings around
the Centaur-object (10199) Chariklo \citep{chariklo},
and possibly Chiron (2060) \citep{chiron1, chiron2},
vividly demonstrates that small solar
system bodies may indeed possess rings in their own right
(also see the discussion in \citealt{sicard}).

Apart from ``fossil'' rings left over from the
initial creation of the satellite system, we might expect to find diffuse
dust rings arising from
ongoing impact erosion of the minor satellites.
\citet{durda} argued that the frequency of collisions
within the Kuiper Belt is high enough to cause significant
erosion of all small KBOs over the age of the solar system.
In the specific context of Pluto, \citet{stern06}, \citet{hst_ring},
\citet{s09}, and \citet{poppe} predicted that
impact gardening of the minor satellites Hydra and Nix
(and by implication Kerberos and Styx, which were discovered
later) could inject fine debris or dust into the environment
around Pluto and Charon, leading to transitory or long-lived
rings. (Escape velocities are too high
for this mechanism to generate rings directly from Pluto or Charon.)

On theoretical grounds, \citet{stern06} estimated that a ring generated from steady-state
erosion of Hydra and Nix would have optical depth $\tau=5\times 10^{-6}.$
\citet{poppe} argued for a shorter particle lifetime, yielding $\tau=10^{-7}.$
Given the dynamic complexity
of Pluto and  Charon plus the four minor satellites,
a key theoretical problem is identifying orbits that can host stable rings.
\citet{poppe} and \citet{solar} showed, for example, that long-lived
rings could exist between the orbits of Nix and Hydra,
as well as at co-orbital locations in the orbits of the four minor satellites.
However, the subsequent discovery of Kerberos orbiting between these two moons
partially invalidated their conclusions \citep{ps15}.

On the other hand, \citet{solar} argued that
solar radiation pressure is sufficient, even at the distance of Pluto, to clear
small particles from the system, given Pluto's overall shallow gravity well.
They predicted that the optical depth of any long-lived
dust ring would be no more than $4\times 10^{-11},$
well below any feasible detection limits.
This result holds despite the substantial rates of impact gardening
estimated by \citet{durda}
(a significantly lower rate may be implied by the crater counts
measured by \citealt{singer}).
Furthermore, simulations by \citet{youdin} and \citet{sh15} show the system
to be chaotic, raising questions about the long-term stability of the small
moons themeselves, irrespective of any embedded rings.
Interior to the orbit of Charon, long-lived rings are unlikely to exist
due to the drag induced by Pluto's extended atmosphere \citep{porter}.

Because this debate has been inconclusive, it remained an open question
whether Pluto might have faint rings above the sensitivity threshold of
the New Horizons (NH) cameras.
Beyond the scientific interest in such rings, their possible existence also
raised concerns about a potential hazard to the spacecraft during its 
passage through the system.

\subsection{Previous Searches for Plutonian Rings}
A few attempts were made to detect rings in advance of the NH encounter.
\citet{hst_ring} used high-resolution images
of Pluto and Charon taken with the ACS/HRC instrument
on board the Hubble Space Telescope (HST) to derive upper limits for the
visibility of any rings in back-scattered, visual-band sunlight.
Their detection limits were controlled by the scattered-light background
and thus varied with projected distance from Pluto.
At the orbit of Hydra, they limited the rings' normal 
$I/F$ to $\sim2.5\times10^{-7}$.
That limit approximately doubles at the inner limit of orbital stability,
which is $\sim42$,000~km.

Like \citet{hst_ring}, we measure intensity $I$ by the dimensionless ratio $I/F$,
where $\pi F$ is the incoming solar optical flux density at Pluto.
With this definition, $I/F$ would equal unity for a
perfectly diffusing ``Lambert'' surface 
when illuminated and viewed normal to the sunlight.
For a body with geometric albedo $A$ viewed at phase angle $\alpha$, its
surface reflectivity would be
$I/F = A~P(\alpha)/P(0)$,
where $P$ is the phase function.

Above, we quote values of ``normal'' $I/F\equiv\mu I/F$,
which is a more useful quantity for describing optically thin rings.
Here the observed $I/F$ is scaled by a factor
$\mu = {\rm cos}(e)$, where $e$ is the emission angle measured from the ring
plane normal.
The $\mu$ factor compensates for the apparent brightening of an optically thin
ring when viewed closer
to the ring plane, so normal $I/F$ describes the
reflectivity that would be detected if $e=0$.
Normal $I/F$ and optical depth $\tau$ are related via
\begin{equation}
\mu I/F = \tau A \times P(\alpha) / P(0) .
\end{equation}
For \citet{hst_ring}, $A$ was extremely
uncertain, plausibly ranging from 0.04 (for bodies resembling dark KBOs) to
$\sim0.38$, which is the geometric albedo of Charon.
Today, we know that any ring dust is most likely to have albedos comparable
to that of the nearby moons, for which $A=0.5$--0.9 \citep{weaver},
higher than previously suspected.
Nevertheless, because $I/F$ is the measured quantity in image analysis,
we discuss most of our findings below in terms of normal $I/F$.
We revisit the optical depth values in the final discussion and summary section.

Other techniques have also been applied to searches for rings of Pluto.
\citet{boissel} and \citet{throop} separately used stellar occultations
to search for rings. Occultations could potentially
detect compact or narrow rings with widths well below the HST
resolution limit. Unlike image measurements, occultations obtain $\tau$
directly;
however, these experiments did not provide better limits for diffuse rings.
Lastly, \citet{herschel} used far-IR images ($70~\mu{\rm m}$)
from the Herschel Telescope
to look directly for thermal emission from dust around Pluto and Charon,
but again found limits broadly compatible with the HST
results of \citet{hst_ring}.

\subsection{Overview}

The initial summary of science results from the NH encounter
noted that no dust features or rings with normal $I/F>10^{-7}$
were detected based on a preliminary analysis of the
imaging searches conducted by the spacecraft during its approach \citep{nh}.
We have three goals for this dedicated ring-search paper
that go beyond this initial result.
The first is to summarize the results of a preliminary ring search performed
using HST during 2011 and 2012.
The second is to refine the limits derived from the initial analysis of NH data,
as well as to present new limits based on data sets that had not yet been
downlinked when \citet{nh} was published.
These include most of the deepest imaging obtained by the NH cameras.
The third is to describe the NH search observations
and associated analysis methodologies.

\section{The HST Ring Search}
\label{sec:hst_back}

\begin{figure}[htbp]
\centering
\includegraphics[keepaspectratio,width=6 in]{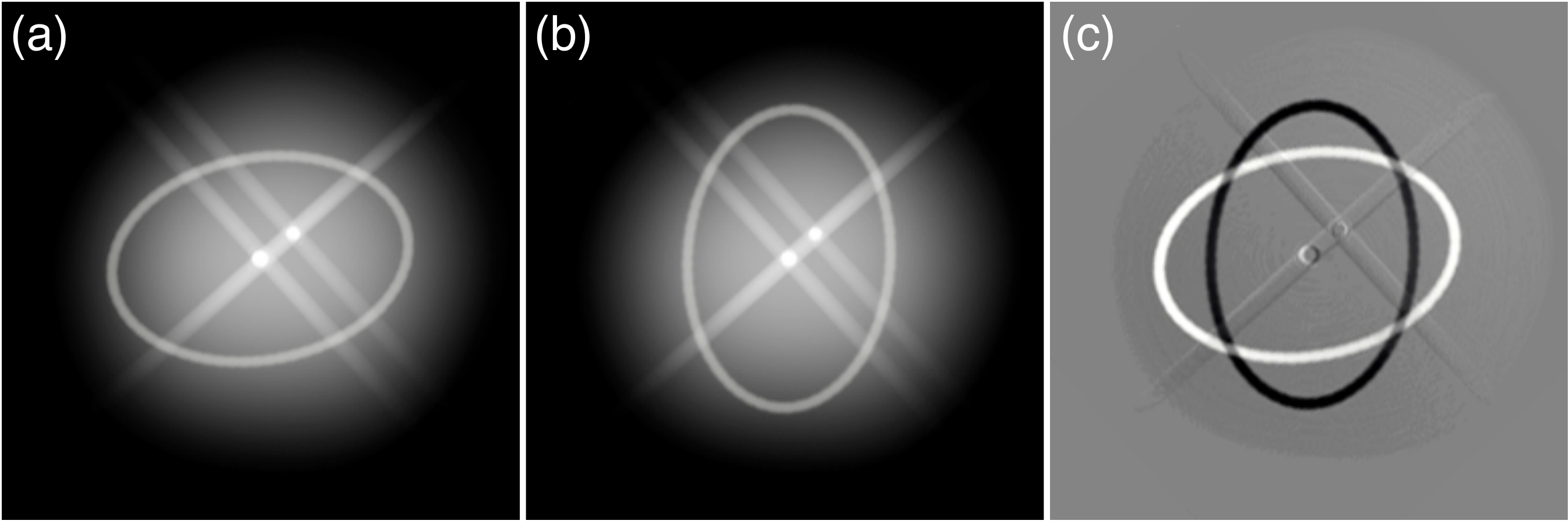}
\caption{The HST imaging strategy employed to search for rings in 2011 and 2012.
Panels (a) and (b) are cartoons of Pluto and Charon surrounded by a glare
pattern plus a hypothetical ring. The camera has been rotated $90^\circ$
between the two exposures, so the glare pattern is similar but the rings
have rotated. Panel (c) shows the difference between these two images.
The glare pattern is nearly canceled out, but positive and negative images of
the ring remain.}
\label{fig:rotation_trick}
\end{figure}

In 2011, coauthors Showalter and Hamilton used HST to search for the putative
rings of Pluto (HST-GO-12346).
The prior search by \citet{hst_ring} was limited by the extensive
glare surrounding Pluto and Charon.
Showalter and Hamilton employed a ``trick'' to model and subtract this glare,
potentially revealing rings $\sim30$ times fainter than the previous upper limit.
Their approach was to image the Pluto system during two HST visits that
differed by a $90^\circ$ rotation of the camera.
The visits were also timed to place Charon at roughly the same position
relative to Pluto on the CCD (Fig.~\ref{fig:rotation_trick}).
The expectation was that that the glare patterns around Pluto and Charon,
which are instrumental in origin, would be nearly identical in the two visits,
but the rings (which are ellipses as projected onto the sky) would be rotated.
By aligning and subtracting the images, most of the planetary glare would cancel out,
leaving behind the signature of a ring as pair of crossed, positive and
negative ellipses.

HST is constrained to keep its solar panels oriented toward the sun, so
rotations $>10^\circ$ are not normally permitted.
The exception is during a brief period around Pluto's opposition, when the
spacecraft can be oriented freely. Table \ref{tab:hstlist}
describes the visits and images obtained near the 2011 opposition.
The ``rotate-and-subtract'' technique worked as planned (see below), although
the processed images did not reveal any rings.
However, because Showalter and Hamilton had obtained longer exposures of the
Pluto system than anyone had previously attempted, these images revealed tiny
Kerberos, Pluto's fourth known moon \citep{iauc2011}, which is $\sim10$\%
as bright as Nix and Hydra.

\begin{deluxetable}{lccccccll}
\tabletypesize{\scriptsize}
\tablecolumns{8}
\tablewidth{0pt}
\tablecaption{Ring Search Sequences with HST}
\tablehead{\colhead{Program}&\colhead{Visit}&\colhead{Date}&\colhead{Orbits}
&\colhead{Filter}&\colhead{$T_{\rm exp}$}&\colhead{$N$}&\colhead{First filename}
&\colhead{Orientation}}
\startdata
GO-12346&21&2011-06-28&2&F606W&479&10&ibo821kvq\_flt.fits&Horizontal\\
GO-12346&22&2011-07-03&2&F606W&479&10&ibo822taq\_flt.fits&Vertical\\
GO-12801&31&2012-06-27&3&F350LP&174&35&ibxa31rmq\_flt.fits&Horizontal\\
GO-12801&32&2012-07-09&3&F350LP&174&35&ibxa32cuq\_flt.fits&Vertical\\
GO-12801&41&2012-06-29&3&F350LP&174&35&ibxa41bwq\_flt.fits&Horizontal\\
GO-12801&43&2012-06-11&3&F350LP&174&35&ibxa43ebq\_flt.fits&Vertical\\
GO-12801&61&2012-06-26&3&F350LP&174&35&ibxa61ifq\_flt.fits&Horizontal\\
GO-12801&62&2012-07-07&3&F350LP&174&35&ibxa62o8q\_flt.fits&Vertical\\
\enddata
\tablecomments{$T_{\rm exp}$ is the exposure time in seconds; $N$ is the number
of images taken using this exposure time;
Orientation indicates the approximate direction of the long axis of
the ellipse representing a circular ring projected on the sky.}
\label{tab:hstlist}
\end{deluxetable}

Following the Kerberos discovery, Weaver et al.\ (GO-12801) led a far more
extensive HST observing program during Pluto's 2012 opposition.
The goal was to perform comprehensive search for rings and small moons.
In addition to the scientific interest, these observations supported the
programmatic goal of assessing possible hazards to the NH spacecraft during
the upcoming Pluto flyby.
This program revealed Styx, Pluto's fifth moon \citep{iauc2012}.
Weaver et al.\ repeated the rotation ``trick'' of the previous year
(Fig.~\ref{fig:rotation_trick}), but with more integration time and a wider
filter (Table \ref{tab:hstlist}).
Here we derive the most sensitive Earth-based limits from a combined
analysis of all these images.

\subsection{Analysis Procedures}

\begin{figure}[htbp]
\centering
\includegraphics[keepaspectratio,width=5 in]{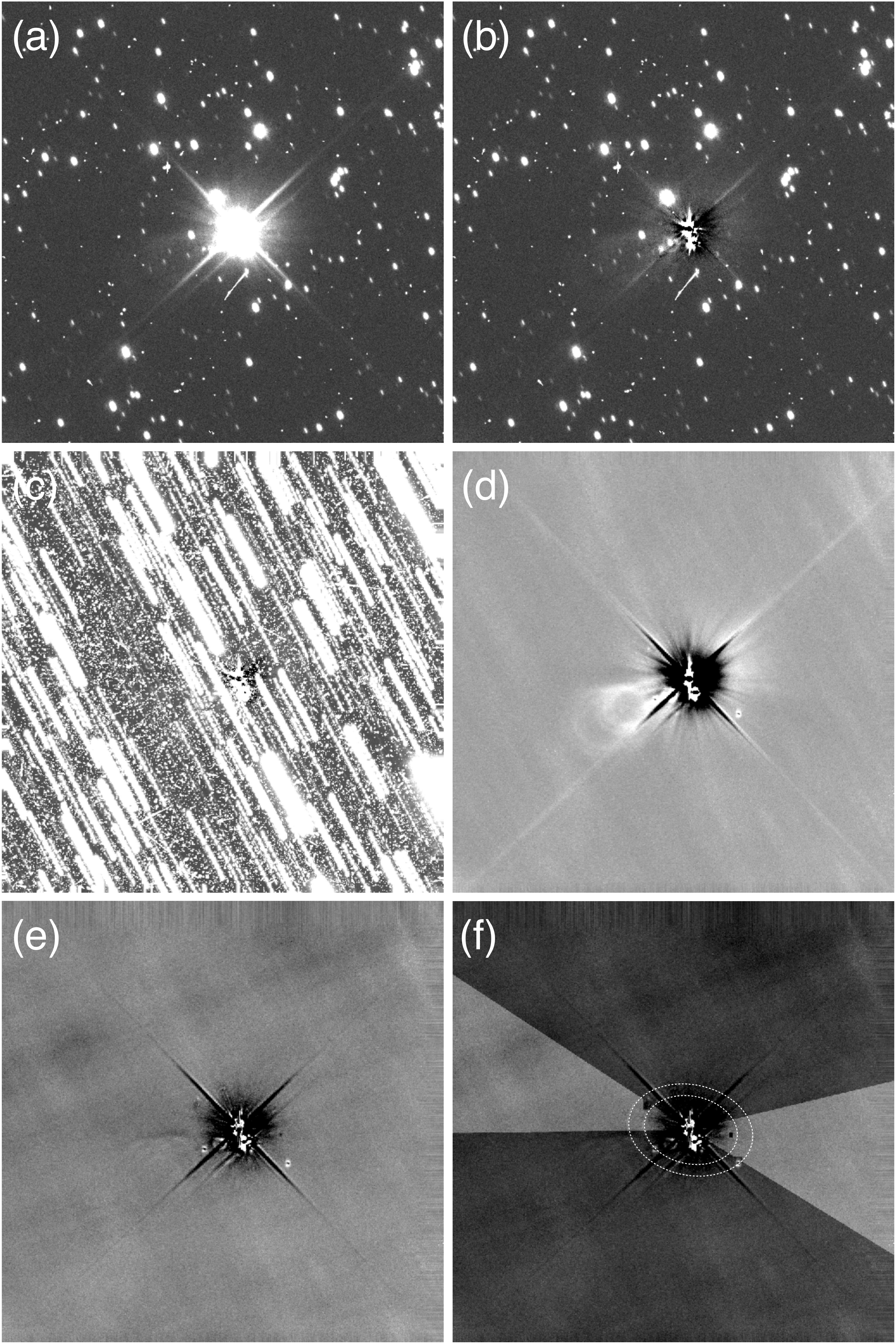}
\caption{Steps in the processing of HST images. (a) The first image from
Visit 31; see Table \ref{tab:hstlist}. (b) The same image after subtraction
of PSF models for Pluto and Charon. (c) Our procedure to stack the images and
sort the pixels at any location by intensity
brings all the stars (and cosmic rays) to the front.  (d) Stripping away the
top layers and coadding the remainder produces a very clean image.
This image has been enhanced by a factor of four relative to the image in
the previous panel.
(e) Subtracting
pairs of perpendicular images would potentially reveal crossed positive and
negative ellipses, but do not. (f) We derive radial profiles of the
possible rings from wedges of the image defined to eliminate diffraction
spikes, known moons, and other residual flaws. Darkened regions indicate
pixels that are excluded. The orbits of Nix and Hydra are shown for reference.}
\label{fig:hst_steps}
\end{figure}

Figure \ref{fig:hst_steps}a shows a typical image from Visit 31 in GO-12801.
Because Pluto is located near the galactic center, numerous background stars
are visible.
Although visits span up to 4 hours, Pluto's moons move only slightly
in this time frame. HST time is measured in units of orbits of HST around the Earth, where
each 95-minute orbit provides about 45 of Pluto viewing.
Small dither steps in the middle of each HST orbit, and also between orbits,
helped to mitigate the effects of hot pixels.
Pluto and Charon saturated during these long exposures, but additional,
short exposures of a few seconds show them clearly and without saturation.
Within each orbit, we navigated the images using the unsaturated images of
Pluto and Charon, plus Nix and Hydra from the longer exposures.
From this information, we derived the pixel coordinates of the system
barycenter within each image.

After the navigation, we subtracted out the known bodies from the images.
We used the program ``Tiny Tim'' \citep{krist} to model the
point-spread function (PSF) for the images.
We then applied a fitting procedure to locate and scale the PSF for each body.
For Pluto and Charon, we also applied a small blur to account for the
fact that these bodies are partially resolved by the instrument.
We masked out the pixels where Pluto and Charon saturated, using only the
unsaturated pixels for this modeling.
The models produced by Tiny Tim are imperfect, but subtracting them
suppresses much of the PSF, including most of the long diffraction spikes that
extend diagonally outward from Pluto and Charon (Fig.\ref{fig:hst_steps}b).

At this stage, the bodies of the Pluto system have been suppressed, but
background stars remain.
A simple coadd of these images would be corrupted by the numerous
stars.
However, the stars are clearly identifiable by their motion between
consecutive images.
We devised a modified ``coadd'' procedure that is much more
successful at suppressing the stars.
We began by aligning all the images from each visit on the
system barycenter and then stacking them into cubes.
Within these cubes, we then sorted the pixels of each ``column'' from
minimum to maximum, where the column is all pixels at the same
pixel offset from the system barycenter in the image layers of the cube.
This caused the background stars, as well as cosmic rays,
to ``rise to the top'' (Fig.~\ref{fig:hst_steps}c).
The final step was to eliminate the top layers and coadd the remainder
(Fig.~\ref{fig:hst_steps}d)

\begin{figure}[htbp]
\centering
\includegraphics[keepaspectratio,width=4 in]{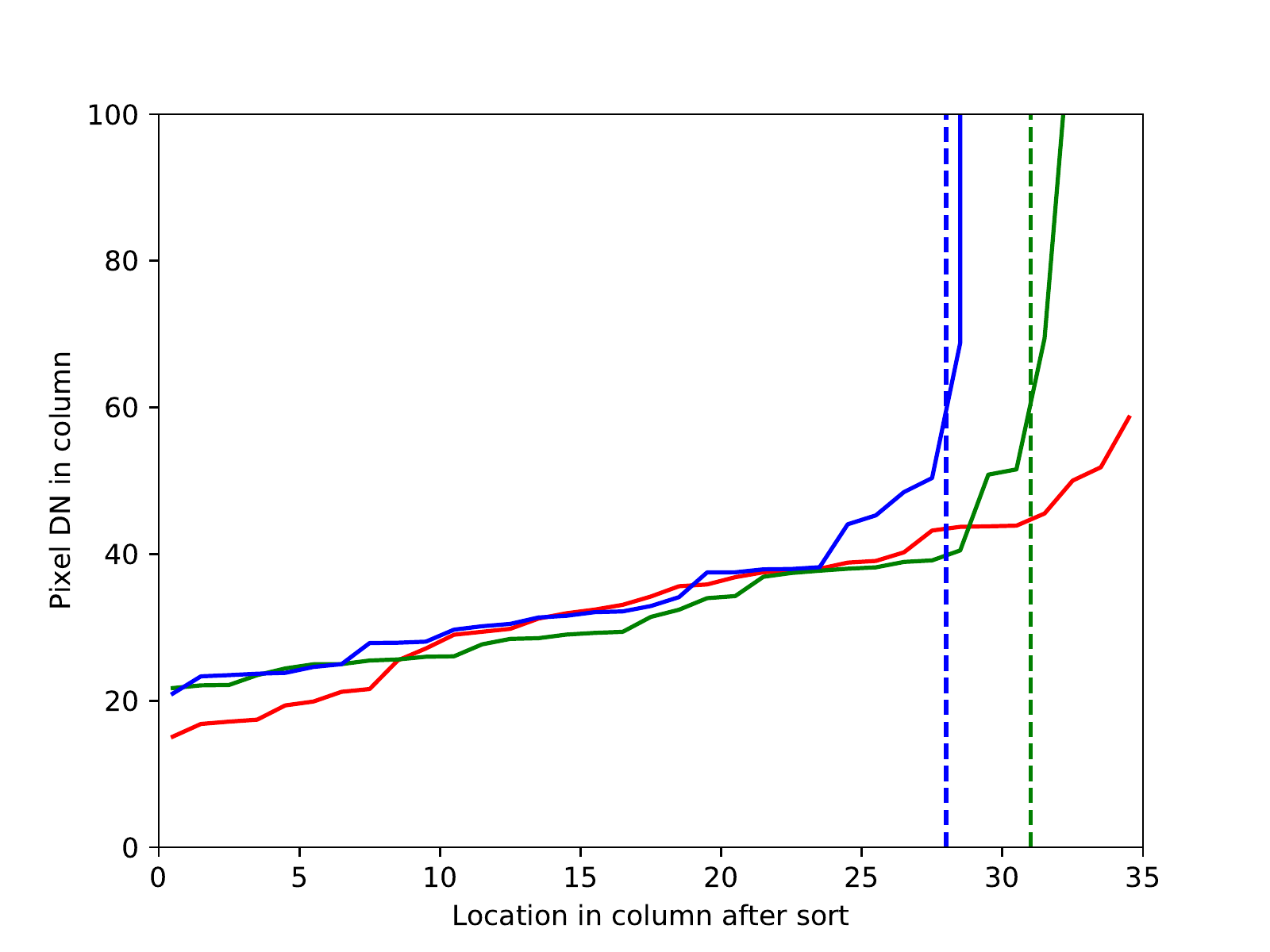}
\caption{We plot DN vs. layer number in a selection of colums from the stack
of 25 deep exposures taken during Visit 31. Near the right, large jumps
upward correspond to background stars. Vertical dashed lines indicate the cutoff
point; layers of the stack above this line are excluded from the coadding used to
produce Fig. \ref{fig:hst_steps}d. The red curve is well-behaved throughout,
indicating that no stars were present at this location so all 35 layers could
be coadded.}
\label{fig:sorted_columns}
\end{figure}

In this processing, the key question was how many of the top layers to remove.
Eliminating a fixed number of layers had the disadvantage that, in order to be
sure all the stars were removed, many perfectly valid pixels were also
excluded from the analysis.
Our more refined procedure took advantage of the fact that, upon examining
the trend in brightness going upward through each column of the stack, it was
easy to identify the point at which a star first appeared
(Fig.~\ref{fig:sorted_columns}).
Working upward through the pixels in each column,
we defined the cutoff point as the first layer at which the value changed by more
than $3\sigma$ relative to the calculated mean and standard deviation of the
pixels below.
Examples of the cutoff point are indicated in (Fig.~\ref{fig:sorted_columns});
the result of this
procedure for Visit 31 is shown in Fig.~\ref{fig:hst_steps}d.

\begin{figure}[htbp]
\centering
\includegraphics[keepaspectratio,width=4 in]{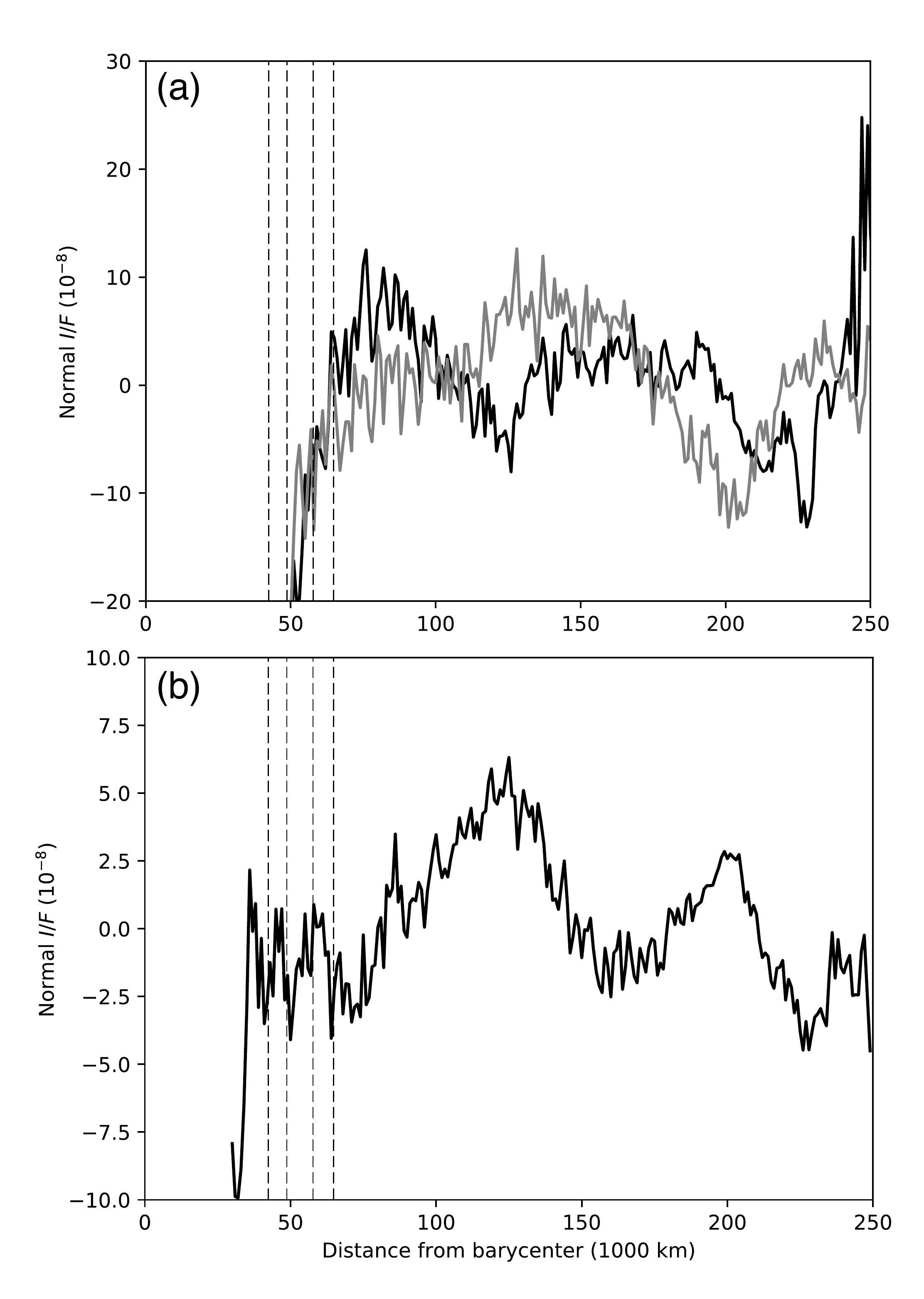}
\caption{(a) Ring profiles derived from Visit 31 (cf.\ Fig.~\ref{fig:hst_steps}f).
Linear trends, which are caused by subtle variations in the background
glare pattern, have been removed. If rings are
present, we would expect these profiles to show an increase at the
same projected radius in each profile. Instead, the variations we see
are uncorrelated and presumably caused by background gradients and residual
stars. Vertical dashed lines indicate the orbits of the four small moons.
(b) Coadding all the finest profiles
produces our best Earth-based opportunity to detect Plutonian rings.
Note the expanded vertical scale here relative to panel (a).}
\label{fig:hst_profiles}
\end{figure}

With the visits now suitably coadded, we aligned the rotated pairs of visits
on the system barycenter and subtracted (Fig.~\ref{fig:hst_steps}e) to potentially
reveal rings as perpendicular, positive and negative ellipses
(cf.\ Fig.~\ref{fig:rotation_trick}c).

We derived two radial profiles of the rings from each HST visit by selecting 
wedges of pixels centered around the long axis of the ellipse that represents
a circular ring projected onto the sky. 
The wedges excluded the residual
diffraction spikes and other obvious image flaws (Fig.~\ref{fig:hst_steps}f).
When a known moon fell within this wedge, we masked it out.
The remaining pixels were sorted into bins defined by the projected
radial distance from the barycenter.
This yielded four profiles from each pair of visits, using horizontal wedges
from ``horizontal'' Visits 21, 31, 41, and 61, plus vertical wedges (reversed
in sign) from their counterparts, Visits 22, 32, 43 and 62.
Although we derived four ring profiles from each set of images,
all profiles were derived from disjoint subsets of the image pixels,
and so they are statistically independent.
Figure \ref{fig:hst_profiles}a shows examples from Visit 31.

It takes several steps to convert HST's pixel values to $I/F$. We work with
calibrated images, whose filenames end with ``\_flt''.
Within these files, the numeric value of each pixel, sometimes referred to as
DN, is equal to the number of electrons accumulated in that pixel of the
CCD during the exposure time $T_{\rm exp}$.
The value of $I/F$ can be determined from
\begin{equation}
I/F = {\rm DN}~{\rm PHOTFLAM}~\pi R_{\rm AU}^2 / (T_{\rm exp}~d\Omega~F_{\sun})~,
\end{equation}
where $R_{\rm AU}$ is the
Sun-Pluto distance in AU, $d\Omega$ is the solid angle subtended by a pixel, and $F_{\sun}$ is
the solar flux density at 1 AU.
Note that $F_{\sun}$ differs from the ``$F$'' in the denominator of $I/F$
by a factor of $\pi$, as discussed above:
$\pi F\equiv F_{\sun} / R_{\rm AU}^{2}$.
The FITS headers of calibrated images contain a parameter PHOTFLAM, which
scales the conversion from DN to intensity
in physical units: 
${\rm PHOTFLAM}=5.22929\times10^{-20}{\rm~erg~cm^{-2}~s^{-1}~\AA^{-1}~electron^{-1}}$
for F350LP and
$1.15157\times10^{-19}{\rm~erg~cm^{-2}~s^{-1}~\AA^{-1}~electron^{-1}}$
for F606W.
WFC3/UVIS pixels are roughly $0\asec04$ squares, so
$d\Omega\approx3.76\times 10^{-14}$ steradians, although a correction for
camera distortion, defined by the ``pixel area map'', alters this value by
$\sim1\%$.

The solar flux density $F_{\sun}$ must be adapted to the filter
bandpass.
The F350LP bandpass is roughly $0.58\pm0.48~\mu$m, whereas 
the F606W bandpass is $0.59\pm0.22~\mu$m.
To calculate $F_{\sun}$, we use the solar spectrum as tabulated by \citet{bohlin}.
We determine the weighted mean of $F_{\sun}$ based on the filter
and instrument throughput:
$F=4.632\times10^{-2}{\rm~erg~cm^{-2}~s^{-1}~\AA^{-1}}$ for F350LP and
$F=5.379\times10^{-2}{\rm~erg~cm^{-2}~s^{-1}~\AA^{-1}}$ for F606W.
Because these two filters are very wide, this calculation intrinsically assumes
that the rings are neutral in color.
Combining, the final conversion factor from DN to $I/F$ is
$1.2\times10^{-7}$ for
Visits 21 and 22 in 2011; $1.7\times10^{-7}$ for visits 31--62 in 2012.
Conversion to normal $I/F$ requires an additional factor of $\mu={\rm cos}(e)$,
where $e=44\adeg0$ in 2011 and $42\adeg5$ in 2012.

Returning to Fig.~\ref{fig:hst_profiles}a,
if a ring were detectable around Pluto, we would see a reflectivity increase 
at a consistent, repeatable location in these profiles.
We do not.
Instead, the variations among the profiles define our level of sensitivity
to the rings.
Our firmest limit is obtained by combining all the cleanest profiles into one
(Fig.~\ref{fig:hst_profiles}b).
The variations imply that normal $I/F<4\times10^{-8}$ near the small moons and
$<8\times10^{-8}$ for possible distant broad rings.

\section{Searching by Back-Scattered Light using New Horizons}

\subsection{Design of the Observational Program}
\label{sec:back}

We conducted an extensive search for satellites and rings starting nine weeks
prior to the closest approach to Pluto.
This was done as a mission-critical
program to identify material that might be hazardous to the
spacecraft in time to select
a safer flyby trajectory.
In addition to searching directly for unknown rings, 
the program also searched for additional small satellites,
which might be markers of rings too diffuse to be detected
directly in back-scattered light (see \citealt{weaver}).

We employed the Long Range
Reconnaissance Imager (LORRI), a $1024\times1024$
pixel CCD camera mounted at the Cassegrain focus of a 20.8 cm reflector
with a $17\amin4\times17\amin4$ field of view.
The native LORRI pixel scale is $1\asec02/$pixel;
however, as we discuss below, the ring search used $4\times4$
on-chip binning, producing $256\times256$ images with a $4\asec08/$pixel scale.
LORRI is a panchromatic imager covering 350 nm to 850 nm
(these wavelengths mark the band-pass FWHM).
See \citet{lorri} for a full description of the instrument.

Table \ref{tab:obs} lists the complete set of NH observing sequences
specifically designed for ring searches.\footnote{The table
omits an additional inbound series of 432 LORRI $4\times4$ observations
that had been scheduled for to occur eight days before closest approach.
This sequence (denoted O\_SAT\_RING\_DEEP) was never executed
due to an onboard spacecraft anomaly, which halted observations
for approximately two days.}
The New Horizons Project referred to the approach-phase effort as the
``hazard-search program'', so all of these images have a common
``U\_HAZARD'' designation.
The hazard sequences were augmented with deep LORRI imaging
of the search fields taken one and two years before the encounter
to provide a reference background.
These sequences are listed in the first four lines of Table \ref{tab:obs}.
Results obtained by this program provided the normal $I/F<10^{-7}$
limits on rings reported by
\citet{nh} in the initial summary of scientific results from the encounter.

\begin{deluxetable}{llcrrllll}
\tabletypesize{\scriptsize}
\tablecolumns{9}
\tablewidth{0pt}
\tablecaption{Ring Search Sequences}
\tablehead{\colhead{ }&\colhead{ }&\colhead{ }&\colhead{Phase}
&\colhead{ }&\colhead{MET}&\colhead{MET}&\colhead{UT S/C}&\colhead{Encounter}\\
\colhead{Name}&\colhead{Instr.}&\colhead{Type}&\colhead{(Deg)}
&\colhead{N}&\colhead{Start (s)}&\colhead{End (s)}&\colhead{Start}&\colhead{Day}}
\startdata
13182:A7LR090A\_01\_U\_HAZARD\_SIM&LORRI&B&13.3&60&235006328&235008843&182 17:40:05&P$-743$\\
13182:A7LR090B\_01\_U\_HAZARD\_SIM&LORRI&B&13.3&30&235025528&235025818&182 23:00:05&P$-743$\\
14195:A8LR094A\_01\_U\_HAZARD&LORRI&B&14.3&48&267960128&267962895&199 03:30:05&P$-361$\\
14195:A8LR094B\_01\_U\_HAZARD&LORRI&B&14.3&48&268133828&268136595&201 03:45:05&P$-359$\\
U\_HAZARD\_131A&LORRI&B&14.9&48&293653927&293656172&131 12:40:00&P$-64$\\
U\_HAZARD\_131B&LORRI&B&14.9&48&293681167&293683412&131 20:14:00&P$-64$\\
U\_HAZARD\_132&LORRI&B&14.9&48&293763127&293765372&132 19:00:00&P$-63$\\
U\_HAZARD\_149\_150A&LORRI&B&14.9&48&295200607&295202852&149 10:18:01&P$-46$\\
U\_HAZARD\_149\_150B&LORRI&B&14.9&48&295246327&295248572&149 23:00:01&P$-46$\\
U\_HAZARD\_149\_150C&LORRI&B&14.9&48&295262527&295264772&150 03:30:01&P$-45$\\
U\_HAZARD\_156A&LORRI&B&15.0&72&295787767&295790270&156 05:24:01&P$-39$\\
U\_HAZARD\_156B&LORRI&B&15.0&72&295838827&295841330&156 19:35:01&P$-39$\\
U\_HAZARD\_156C&LORRI&B&15.0&72&295849927&295852430&156 22:40:01&P$-39$\\
U\_HAZARD\_166&LORRI&B&15.0&96&296651467&296654214&166 05:19:01&P$-29$\\
U\_HAZARD\_167A&LORRI&B&15.0&96&296720287&296723034&167 00:26:01&P$-28$\\
U\_HAZARD\_167B&LORRI&B&15.0&96&296782147&296784894&167 17:37:01&P$-28$\\
U\_HAZARD\_167C&LORRI&B&15.0&96&296793727&296796474&167 20:50:01&P$-28$\\
U\_HAZARD\_173&LORRI&B&15.0&96&297294127&297296874&173 15:50:01&P$-22$\\
U\_HAZARD\_174A&LORRI&B&15.0&24&297383557&297384215&174 16:40:31&P$-21$\\
U\_HAZARD\_174B&LORRI&B&15.0&24&297391450&297392108&174 18:52:04&P$-21$\\
U\_HAZARD\_177A&LORRI&B&15.0&24&297613597&297614255&177 08:34:31&P$-18$\\
U\_HAZARD\_177B&LORRI&B&15.0&24&297620060&297620628&177 10:22:14&P$-18$\\
U\_HAZARD\_182A&LORRI&B&15.0&24&298073888&298074741&182 16:26:01&P$-13$\\
U\_HAZARD\_182B&LORRI&B&15.0&24&298078918&298079918&182 17:49:51&P$-13$\\
U\_SAT\_RING\_L4X4\_1&LORRI&B&15.2&48&298855628&298856348&191 17:35:00&P$-4$\\
U\_SAT\_RING\_L4X4\_2&LORRI&B&15.6&24&299078218&299078698&194 07:24:50&P$-1$\\
Alice\_StarOcc1&Alice&S&168.8&691&299195100&299204100&195 15:53:00&P$+4$h\\
U\_TBD\_4&MVIC&F&166.0&4&299234946&299234376&196 02:57:07&P$+1$\\
O\_RingDep\_A\_1&MVIC&F&165.5&17&299291206&299292406&196 18:34:47&P$+1$\\
O\_RING\_OC2&Alice&S&165.3&485&299345347&299348940&196 09:37:01&P$+1$\\
O\_RING\_OC3&Alice&S&165.3&475&299387827&299391420&196 21:25:01&P$+1$\\
O\_RING\_DEP\_LORRI\_202&LORRI&F&165.0&300&299789519&299791473&202 13:00:00&P$+7$\\
O\_RING\_DEP\_MVICFRAME\_202&MVIC&F&165.0&14&299793105&299796965&202 13:59:47&P$+7$\\
O\_RING\_DEP\_LORRI\_211&LORRI&F&165.0&300&300581519&300583474&211 17:00:00&P$+16$\\
O\_RING\_DEP\_MVICFRAME\_211&MVIC&F&165.0&14&300584506&300588366&211 17:49:47&P$+16$\\
O\_RING\_DEP\_LORRI\_305A&LORRI&F&164.8&300&308703899&308705894&305 17:13:00&P$+110$\\
O\_RING\_DEP\_MVICFRAME\_305A&MVIC&F&164.8&14&308710566&308706706&305 17:59:47&P$+110$\\
O\_RING\_DEP\_LORRI\_305B&LORRI&F&164.8&300&308711819&308714626&305 19:25:00&P$+110$\\
O\_RING\_DEP\_MVICFRAME\_305B&MVIC&F&164.8&14&308718486&308714626&305 20:11:47&P$+110$\\
O\_RING\_DEP\_LORRI\_305C&LORRI&F&164.8&300&308719739&308721734&305 21:37:00&P$+110$\\
O\_RING\_DEP\_MVICFRAME\_305C&MVIC&F&164.8&14&308726406&308722546&305 22:23:47&P$+110$\\
\enddata
\tablecomments{ Type specifies the detection methodology as `B' for
backscattered light, `S' for stellar occultation,
and `F' for forward scattering.  N is number of image files.
LORRI files contain single exposures; MVIC files contain multiple exposures.
Each Alice file contains roughly 7-12 seconds of continuous data.
MET is the mission elapsed time measured in seconds from the launch
of New Horizons; it provides unique observation IDs.
The last two columns give sequence start times in UT (S/C = spacecraft frame),
and in days (or hours when marked) relative to Pluto closest approach.
The leading figure in the UT column is the numbered day of the year,
which is 2015 but for the first 4 entries.}
\label{tab:obs}
\end{deluxetable}

\begin{figure}[htbp]
\centering
\includegraphics[keepaspectratio,width=4 in]{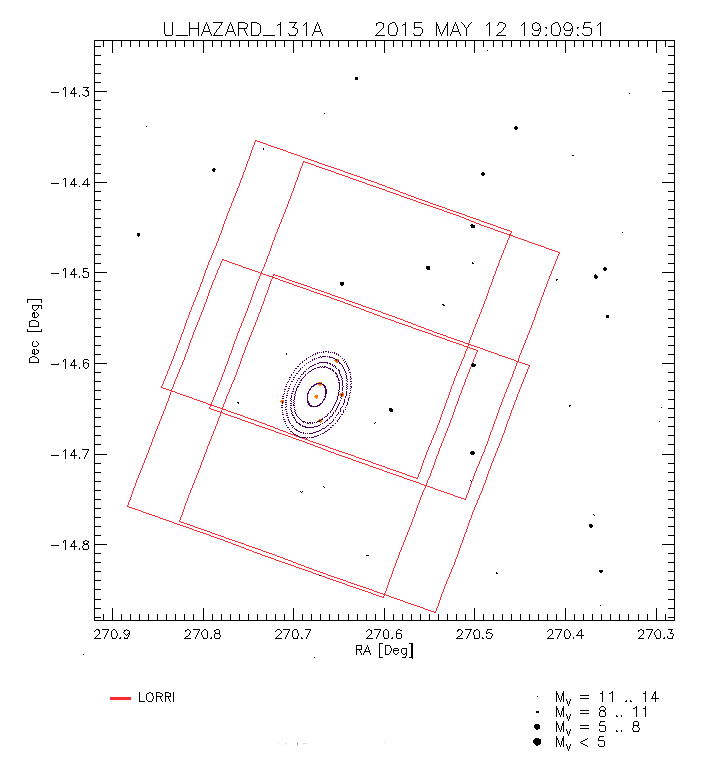}
\caption{The footprint on the sky typical for most of the
$2\times2$ LORRI mosaics used for back-scattered light searches is shown.
This example is from the \texttt{U\_HAZARD\_131A} sequence,
which occured at P$-64.$
Note that the mosaic is aligned with and elongated along the projected
major-axis of the orbits of the minor satellites.  There is a small-scale
random pattern in the relative positioning of the four LORRI fields
due to accuracy limits in the absolute positioning of the spacecraft.}
\label{fig:hazard_foot}
\end{figure}

\begin{figure}[htbp]
\centering
\includegraphics[keepaspectratio,width=4 in]{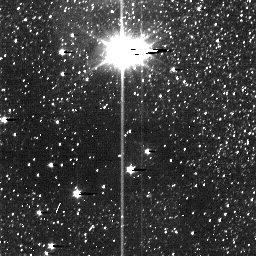}
\caption{The first $4\times4$ LORRI image, lor\_0296651467, in the P$-29$
hazard-search sequence is shown to exhibit the typical features present
in the search imagery.  The exposure is 10s.  Note the heavily crowded
background from the Milky Way bulge.  Hot pixels and cosmic-ray events
are also evident.  Pluto and Charon are both heavily overexposed.  The
vertical streaks spanning the full frame are charge left over from
an incomplete frame-transfer correction due to the saturated centers
of Pluto and Charon.  Both objects, as well as the brightest stars, also
exhibit ``amplifier undershoot" tails to the right of the core of the images,
which pull the local signal below the sky level.}
\label{fig:hazard_4x4}
\end{figure}

The hazard search comprised seven epochs
of observations, each divided into two or three sequences
or sub-epochs of imaging, separated by a few hours to a day, to 
enable the detection of satellites by their orbital motion.
In each sub-epoch, a $2\times2$ mosaic of heavily-overlapping
LORRI fields covered the region around Pluto (Figure \ref{fig:hazard_foot}).
We set the orientation to align with the projected major-axis
of the satellite system in the CCD y-direction.
At each position in the mosaic, we obtained multiple 10s exposures.
For most of the epochs, we obtained images at each position
using two roll angles of the spacecraft separated by $90^\circ$.
This made it possible to 
mitigate the obscuration of portions of the field by the strong diffraction
spikes, incompletely corrected charge-transfer smearing in the column direction,
and ``amplifier undershoot'' trails in the row direction
produced by the disks of Pluto
and Charon, which were severely over-exposed in all sequences.
The detailed number of exposures, as well as the temporal sampling intervals,
varied over the hazard sequences.
The deepest search occurred at P$-29$ (29 days prior to closest approach
to Pluto) with 24 images at each of the four mosaic positions for each of
four sub-epochs.\footnote{Throughout the paper we use designations
like P$-29$ or P$+110$ to denote the number of days before ($-$) or after ($+$)
the time of closest approach to Pluto.}

\subsection{Reduction and Analysis of the Observations}

\begin{figure}[htbp]
\centering
\includegraphics[keepaspectratio,width=4 in]{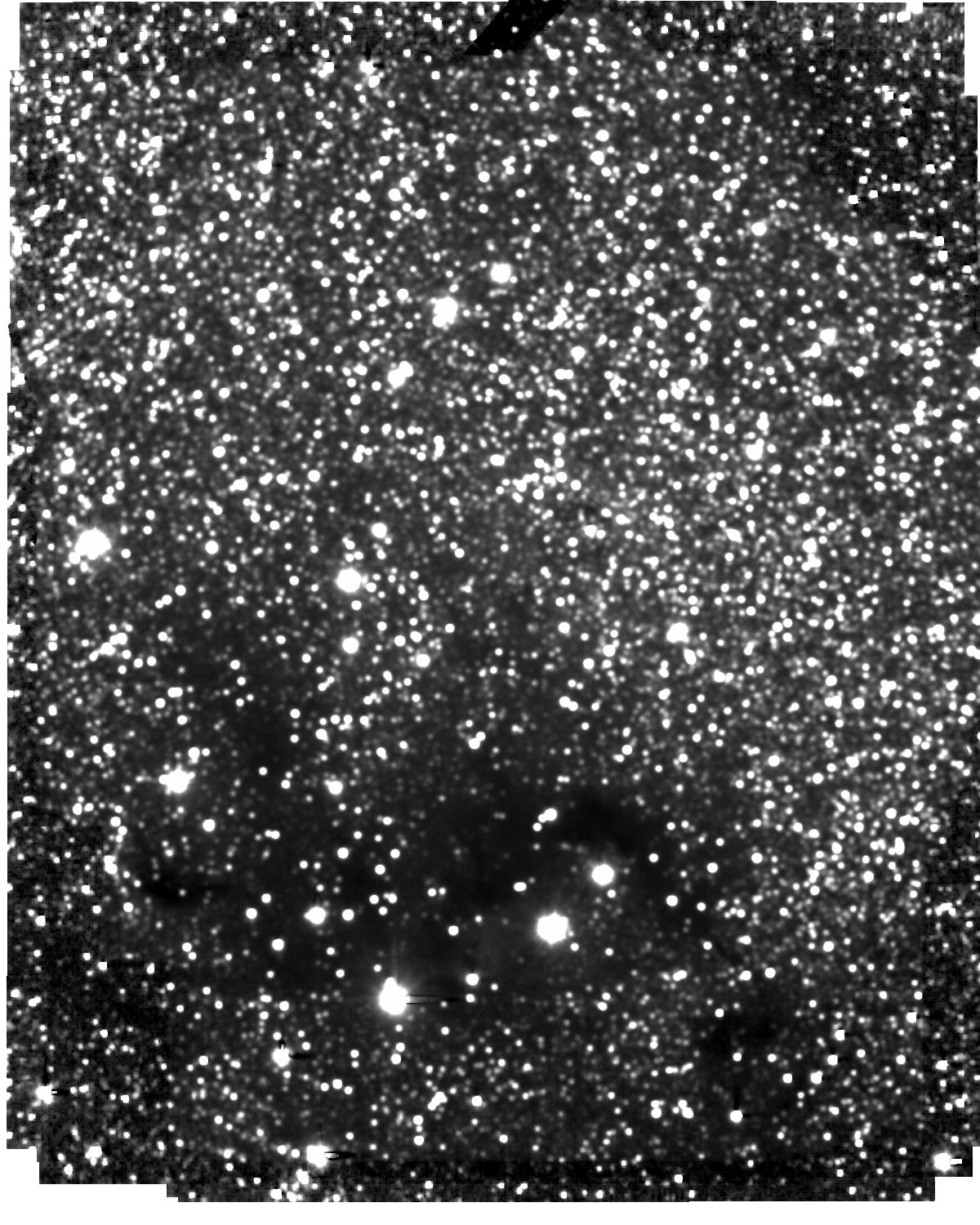}
\caption{For reference we show a deep stack of 340 LORRI $4\times4$
images covering the complete hazard-search area.  The images were
drawn from the sequences in Table \ref{tab:obs} with encounter dates P$-63$
and earlier.  Pluto was present in the field, but was masked out using
its proper motion over the time span of the image set.}
\label{fig:hazard_stars}
\end{figure}

Reduction of the hazard images to effect the detection of faint
satellites and rings presented a strong technical challenge
to the NH team.  During the approach phase, Pluto
was seen in projection against the bulge of the Milky Way,
thus the background was richly crowded with stars.
The images were $4\times4$ binned to 
achieve maximum photometric sensitivity to point sources
(as well as greatly reducing the downlink data volume),
but as a result were severely under-sampled.
The NH pointing system
was optimized for the short integrations
needed during the closest phases of the encounter;
because the spacecraft does not have reaction wheels
for fine pointing control,
the spacecraft attitude typically drifted slightly during the 10s exposures,
though drift was minimized by firing thrusters to maintain pointing
within a narrow (arcsecond-class) deadband.
As such, the PSF of the images could vary markedly from exposure to exposure.
Lastly, the images were also strongly affected by cosmic-ray
events.  We show a typical hazard-search LORRI $4\times4$ 10s
image in Figure \ref{fig:hazard_4x4}.
Note the heavily crowded background.  In a
single image the completeness limit is about $m_V\sim18,$
going to $m_V\sim20$ in the complete stack at any epoch.
A deep image of the full search area is shown in Figure \ref{fig:hazard_stars}.

\begin{figure}[htbp]
\centering
\includegraphics[keepaspectratio,width=4 in]{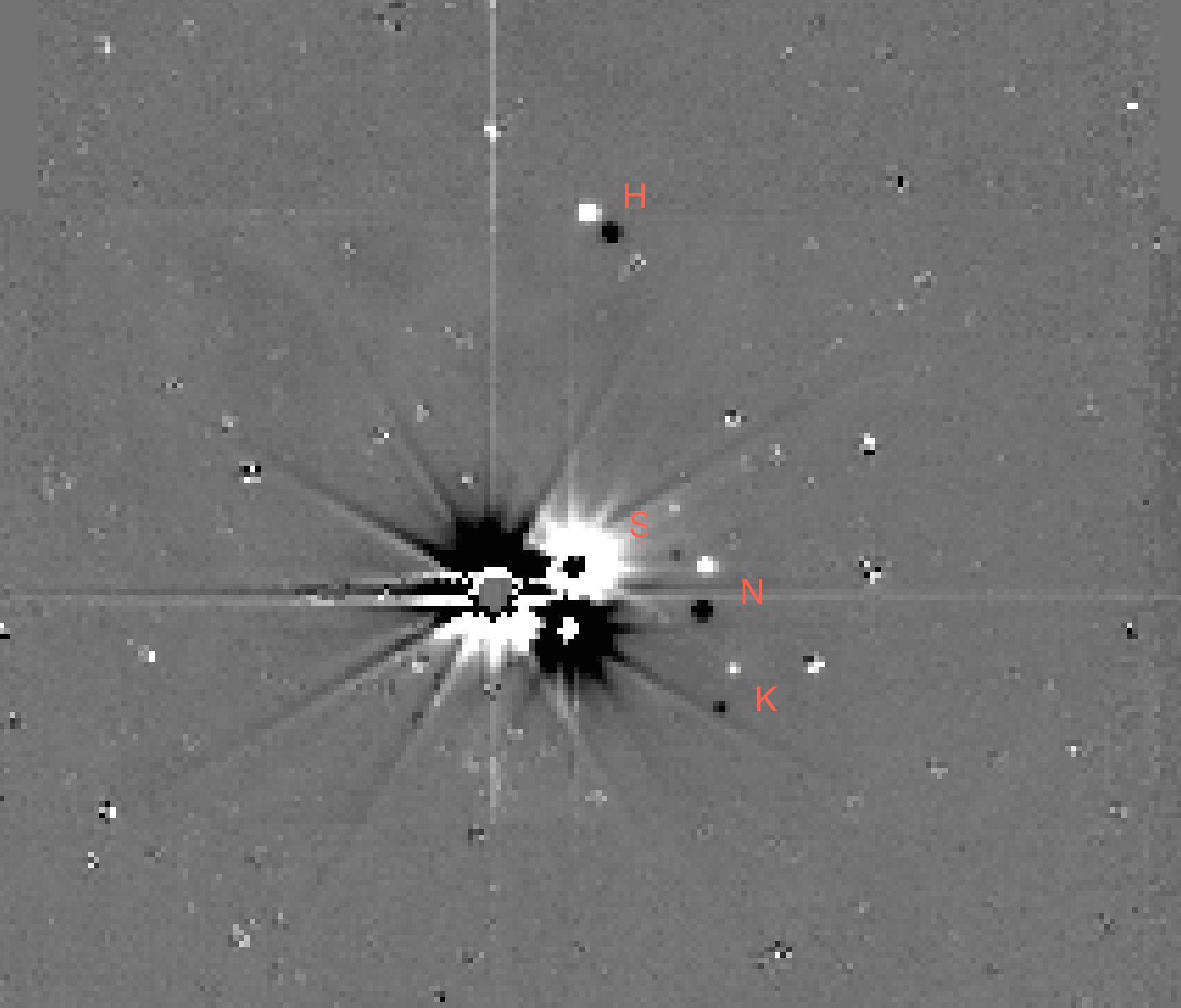}
\caption{A demonstration of the search for faint satellites with
the P$-39$ hazard sequences.  After subtracting the stellar background
using the principal components analysis method stacks are made
at two P$-39$ sub-epochs
separated by 17 hours, which are subtracted form each other.
The known satellites, Hydra, Nix, Styx, and Kerberos (indicated by their
initials) are readily recognized by their motion between the two sub-epochs.
The strong residuals at the center are due the the motion of Pluto (L)
and Charon (R) with respect to the barycenter as well.}
\label{fig:p39_diff}
\end{figure}

Eight members of the NH team examined all of the hazard images.
The experience of this team was considerable and diverse,
encompassing work on crowded stellar fields in
ground-based, HST-based, planetary, and even extra-galactic contexts.
As such, the team deployed a diversity of analysis approaches.
Broadly speaking, the search task could be broken down into three parts.
The first step typically focused on a way to model
or remove the strongly-crowded,
but largely invariant stellar background of the images.
The resulting difference images at a common
sub-epoch, rotation, and position within the larger search field, would
then be aligned and stacked using various statistical procedures to eliminate
left-over residuals from bright stars, cosmic rays, hot pixels, and so on.
Lastly, stacked images at different
sub-epochs would be compared to search for faint satellites or rings
(see Figure \ref{fig:p39_diff}).
The central concern in all approaches was to take
care to preserve large-scale diffuse features indicative of rings,
or transient stellar point sources indicative of possible unknown satellites.
We briefly describe the various strategies.

\begin{figure}[htbp]
\centering
\includegraphics[keepaspectratio,width=4 in]{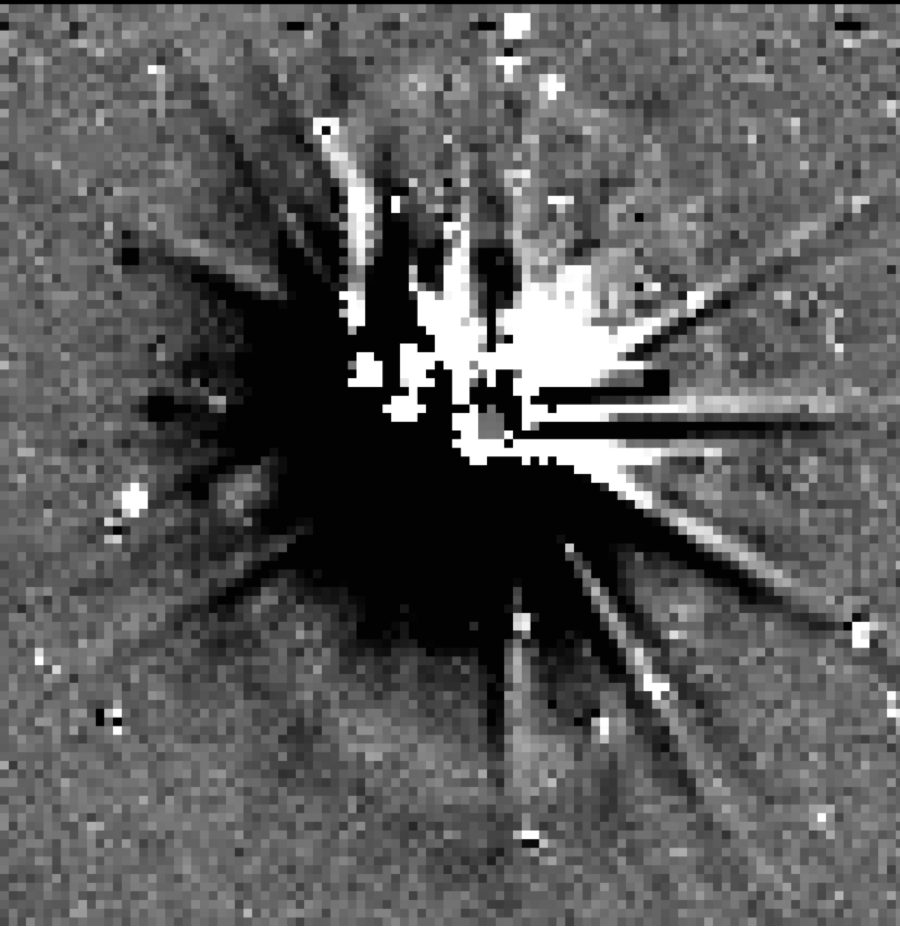}
\caption{Recovery of a simulated ring associated with Styx and having
$I/F=2\times10^{-7}.$ In this test the ring was recovered by differencing
image stacks generated from simulations of the three \texttt{U\_HAZARD\_149}
and three \texttt{U\_HAZARD\_156} sequences obtained on P$-46$ and P$-39$.
The P$-39$ ring image is angularly larger and is subtracted
from the P$-46$ image, creating a difference ring image that is negative
(dark) on the outside and positive (lighter) on the inside.
The simulated image of Styx (positive in P$-46$) is seen embedded in the
ring at the top of the image and slightly right of center.
Pluto and Charon are strongly over-exposed. In this simulation, while
the stellar background was well modeled, no attempt was made to
model Pluto or Charon in this particular case.}
\label{fig:hazard_sim}
\end{figure}

One approach was to characterize the image PSF and pointing variations
at any location with principal components analysis (PCA)
applied to the background reference image sets.
The PCA approach generated a set of ``eigenimages" that
encapsulated the image-to-image variations within the reference set
\citep{pca}.
This allowed a construction of a model reference image
for each image in the actual hazard sequences that attempted
to match the specific PSF and pointing for the image modeled.
Subtracting each model image from each search
image produced a set of difference images
at each search location and epoch that were then stacked to
build up the signal for faint satellites and rings.
This process was done in a few iterative cycles, rebuilding
the initial models in one cycle based on sources rejected in a previous cycle.
Crucially, the stacking included statistical
rejection of cosmic-ray events and other variable defects.
Once the stacks were completed,
rings and faint satellites would be identified by comparing the
stacks made at different epochs.

A second approach used various models of the stellar background based
on stacks either generated from the reference sequences obtained
in advance of the hazard-search, or from other hazard-search epochs
than the one being searched,
depending on whether the goal was to detect satellites versus rings
in the search imagery.  A key step in all cases was to include
models of the light distributions produced by Pluto and Charon, 
including their associated optical ghosts, models of the frame 
transfer smear for the Pluto and Charon images, and a model that reproduces
the ``jailbar'' pattern seen in the raw images.\footnote{The
jailbar pattern was generated by alternating small bias offsets
in the odd versus even columns of the LORRI images, and may be faintly
seen in the example image shown in Figure \ref{fig:hazard_4x4}.}
All the model-subtracted images taken at a single pointing
and roll angle were combined using a robust mean technique to produce a
deep composite image, with most of the original image artifacts removed.

Yet another approach took advantage of the fact that, by random chance,
certain image pairs would occasionally have very similar PSFs. This procedure
cross-correlated every pair of ring search images and, for each image,
and used the ``most similar'' subset of the remaining images to model
and subtract out the star field.
Again, after the subtraction, the resulting
difference images were stacked using robust
statistics to reject cosmic rays and other variable defects.

\begin{figure}[htbp]
\centering
\includegraphics[keepaspectratio,width=4 in]{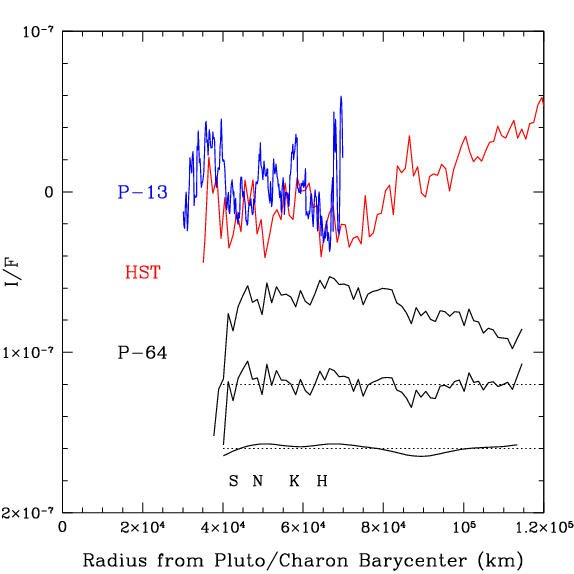}
\caption{Back-scattering $I/F$ radial brightness profiles covering the
the orbital range of Pluto's minor satellites are shown. The orbital
radii of the satellites are indicated by their initials.
The present results measured from LORRI hazard-search images
are labeled with the P$-64$ and P$-13$ sequence labels.
A strong centrally-increasing scattered-light background has been subtracted
from the P$-13$ profile to improve the identification of narrow rings.
``HST'' indicates the final coadded HST I/F profile derived
in the previous section and presented in Figure \ref{fig:hst_profiles}.
The P$-64$ $I/F$ profile has the smoothest background of the
various hazard-search epochs, but has an average level of
$I/F=-7\times10^{-8}.$ Profiles derived by subtracting a parabolic background
from the P$-64$ observed $I/F$ profile; and then smoothing it
to $1.2\times10^4$ km resolution ($8\times$ the native resolution)
are shown below the original P$-64$ profile.}
\label{fig:lorri_back}
\end{figure}

\subsection{Limits on Rings, Debris, and Dust From Back-Scattered Light}

The mission-critical importance of the hazard search engendered
a formal process to demonstrate optimal
analysis strategies well in advance of the encounter.
All search methodologies were tested with extensive high-fidelity
simulations of all the search sequences, which included simulated
rings and satellites that had a wide range of apparent surface brightnesses
and luminosities.
The predicted sensitivity limits on the detection of rings were based
on rings injected into these simulations.
Interestingly, despite the diversity of the analysis techniques,
all reached similar levels of sensitivity.  
As an example, simulated rings close to Pluto with
normal $I/F=2\times10^{-7}$ were
readily identified (Figure \ref{fig:hazard_sim}).
It's clear that fainter rings than this example would still be readily detected.
As discussed below, however, actual $I/F$ detection limits are
determined by direct analysis of the observations;
we did not validate the simulations to serve this purpose.

Instead, we derived upper limits on
normal $I/F$ for the ring intensities
in the actual encounter hazard-search
from our analysis of the
residuals in the deep stacked images at any epoch.
Using the known geometry of the
observations (i.e., the spacecraft location relative to
Pluto and the geometry of the Pluto system),
we created ``backplanes'' for each composite image composed of
($r$, $\theta$) pairs of coordinates for each pixel,
where $r$ is the radial distance of the pixel from
the barycenter of the Pluto system in the satellite orbital plane,
and $\theta$ is the azimuthal angle of the pixel from the reference longitude,
also calculated in the satellite orbital plane.
Points at a fixed distance from the Pluto barycenter form a
foreshortened ellipse in the sky plane (i.e., in the camera image plane).
Assuming that any ring-like structure can be approximated by such a figure,
the sensitivity for ring searches is greatly enhanced by averaging,
or calculating the median of, the pixel intensities over all azimuths
as a function of distance from the barycenter.
Some example radial profiles using these 
azimuthal averages (or medians) are displayed in Figure \ref{fig:lorri_back}.
Although our primary focus was on the region outside the orbit of Charon, we
also sought rings and small moons orbiting interior to Charon.
For that study, the analysis procedure was the same except that the backplane
values of ($r$, $\theta$) were measured from the center of Pluto rather
than from the system barycenter.

\begin{figure}[htbp]
\centering
\includegraphics[keepaspectratio,width=4 in]{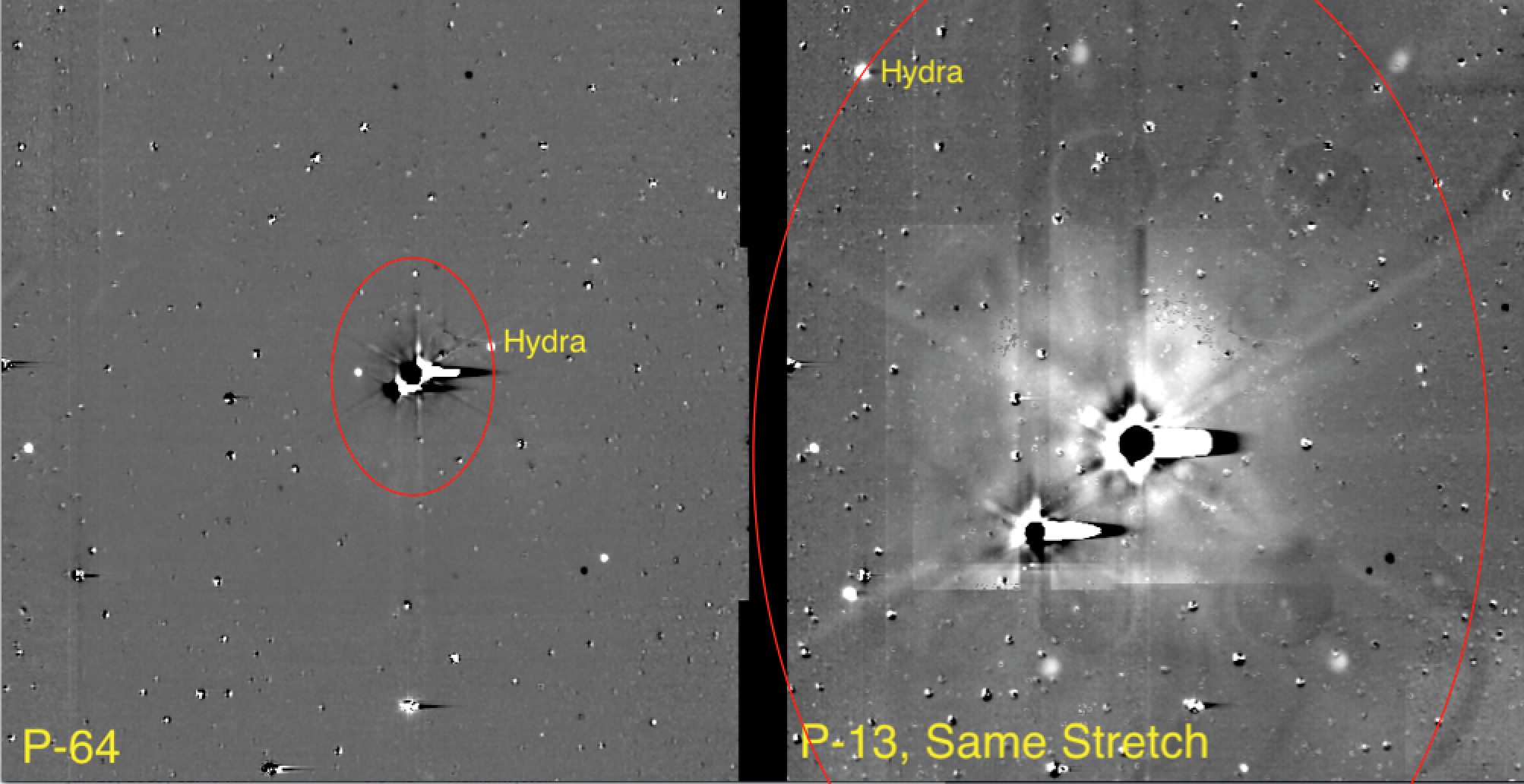}
\caption{Image stacks are shown for the P$-64$ and P$-13$ sequences at the same
stretch to show the increased effects of scattered light from Pluto and Charon
for the later sequences.  The red ellipse indicates the orbit of Hydra.
In each case, non-variable stars have been subtracted
using an image stack taken a year earlier, and models of Pluto
Charon, incorporating large-angle PSFs and optical ghosts,
have been subtracted as well as possible, given model limitations.
This treatment can be contrasted with that in Figure \ref{fig:hazard_sim},
which still includes the residual images of Pluto and Charon.}
\label{fig:lorri_64v13}
\end{figure}

Contrary to our original expectations,
we found that the most sensitive ring search
was derived from the first epoch of observations, which occured at P$-64.$
As the spacecraft drew closer to the Pluto system, the
intensities of the barely resolved Pluto and Charon images grew faster than
the surface brightness of any diffuse (i.e., resolved) structures.
The scattered light and ghosts from the highly saturated Pluto and Charon images
produced bright and spatially complex backgrounds,
which complicated our attempts to detect faint rings or diffuse debris
as the spacecraft moved closer to Pluto (see Figure \ref{fig:lorri_64v13}).

For the azimuthally averaged spatial brightness profiles, we converted
from instrumental units to $I/F$, a commonly-used surface brightness
parmeterization, at the LORRI pivot wavelength \mbox{(6076~\AA)} using:

\begin{equation}
I/F = C / \texttt{RSOLAR} \times \pi R_{\rm AU}^{2} / F_{\sun}
\end{equation}
where $C$ is the count rate in DN/s,
$\texttt{RSOLAR} = 4.092\times10^{6}~{\rm DN/s}~({\rm erg~cm^{-2}~s^{-1}~\AA^{-1}~sr^{-1}})^{-1}$
is the relevant LORRI photometry calibration constant, 
and $F_{\sun}$ is the solar flux, 
\mbox{176 erg cm$^{-2}$ s$^{-1}$ ${\rm\AA^{-1}}$}
at the LORRI pivot wavelength at 1 AU.
To convert to normal $I/F$, we used New Horizons' nearly fixed
emission angle $e=43\adeg2$

A radial brightness profile from the first epoch of the P$-64$ hazard-search
is shown in Figure \ref{fig:lorri_back}.
Scattered light strongly affects the profile inside Styx's orbit
($r_S=42$,413~km), 
but the profile is essentially constant at \mbox{$\sim$~$-7 \times 10^{-8}$}
for larger radial distances covering the orbits of the minor
satellites and extending out to twice the orbit of Hydra.
The residual $I/F$ values are negative because the background for the reference
template image (obtained a year earlier) is systematically larger than
the background for the P$-64$ observations.\footnote{The reason for this is
not understood.  Reference images taken two years in advance
have a {\it lower} background.}
We  conclude that this constant offset of $-7 \times 10^{-8}$
represents a \emph{conservative} upper limit
for any large scale distribution of dust or debris that
would represent a constant background of the scale of the
LORRI field at P$-64$ ($\sim10^5$ km).

Narrow or even broad rings of finer scale, however, should be evident
as compact positive excursions in the $I/F$ profile.
To establish limits on such features, we fitted and subtracted a
smooth background from the P$-64$ profile for
$4\times10^4~{\rm km}<r<1.2\times 10^5~{\rm km}$,
and measured the mean residuals on different spatial scales.
The background-subtracted P$-64$ profile, which now has zero mean,
is plotted below the input P$-64$ profile in Figure \ref{fig:lorri_back}.
At the $1.5\times10^3~{\rm km}$ resolution limit of the profile,
the $3\sigma$ limit on any detectable ring is $I/F\sim2\times10^{-8}.$
This decreases to $I/F\sim1\times10^{-8}$ on $6\times10^3~{\rm km}$
scales and $I/F\sim7\times10^{-9}$ on $1.2\times10^4~{\rm km}$ scales.
The background-subtracted P$-64$ profile, smoothed to $1.2\times10^4~{\rm km}$
resolution is shown as the bottom trace in Figure \ref{fig:lorri_back}.
These limits are $\sim4\times$ fainter than the HST limit over the same region.

The spatial profiles on the later dates have much higher scattered light levels,
as illustrated for P$-13$ in Figure \ref{fig:lorri_back}.
The strongly sloping background and high absolute brightness levels makes
these images much less sensitive to rings or debris.
Even after fitting a smooth curve to the sloping background, the residuals
are much larger for P$-13$ versus P$-64$ (Figure \ref{fig:lorri_64v13}).

\section{Searching by Stellar Occulations}

Near the time of closest approach we also searched for rings, debris, and
dust clouds by the occultation of starlight using
the Alice instrument,
New Horizons' ultraviolet (UV) imaging spectrometer \citep{scs08}.
These sequences are detailed in Table \ref{tab:obs}.
Alice was ideal for stellar occultations of UV-bright stars
because of its ability to take sustained high-cadence observations.
Its primary aperture (the `Airglow' entrance) admits light through a
40~mm diameter aperture.
The projected field-of-view (FOV) is in the shape of a `square
lollipop', with a $2^\circ\times2^\circ$ square adjoining a $4^\circ
\times0\adeg1$ rectangle.
The focal plane has 32 rows (spatial) $\times$ 1024 columns (spectral).
Alice's data is delivered as a list of individual
photons, each tagged with a row, column, and time.
The instrument observes continually, with negligible readout time.

We used Alice for four occultation searches during the Pluto encounter.
The idea of these observations was to search for ring dust or orbital debris
along the line-of-sight between NH
and a distant star.
The theoretical resolution limit is set by the Fresnel limit,
$r_F = \sqrt{D \lambda/2},$
which gives the scale of the smallest resolvable feature in an occultation,
where $D$ is the distance to
the occulting body, and $\lambda$ is the effective wavelength,
which for Alice is $\sim150$~nm.

For each observation, Alice
pointed constantly at a fixed RA/Dec such
that the bright target stars were kept centered on one Alice pixel
during the entire sequence.
As NH traveled on its trajectory, portions of the Pluto
system passed between the stars and the spacecraft. Alice operated continually,
allowing the line-of-sight optical depth to be measured for the duration of the
occultations.

\begin{figure}[htbp]
\centering
\includegraphics[keepaspectratio,width=4 in]{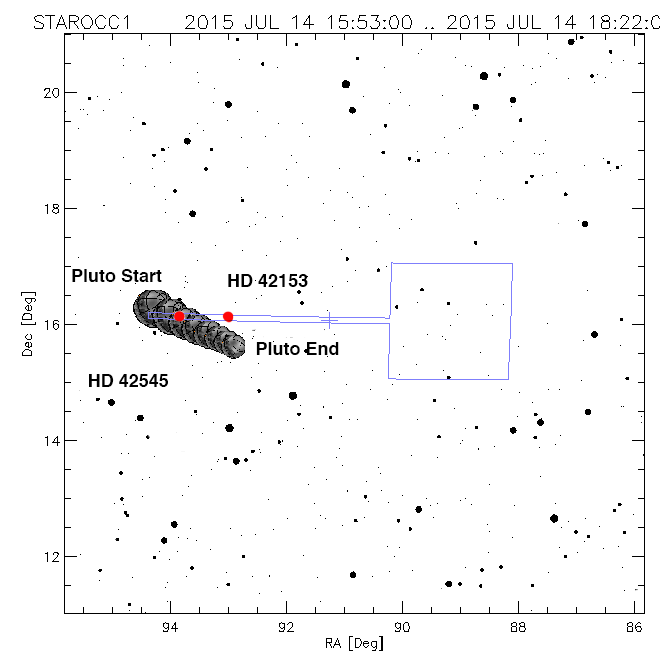}
\caption{Geometry of Alice occultation in the Pluto-Charon, for the
\texttt{Alice\_StarOcc1} sequence.  The blue outline represents the Alice slit.}
\label{fig:geometry_alice_inner}
\end{figure}

\begin{figure}
\centering
\includegraphics[keepaspectratio,width=4 in]{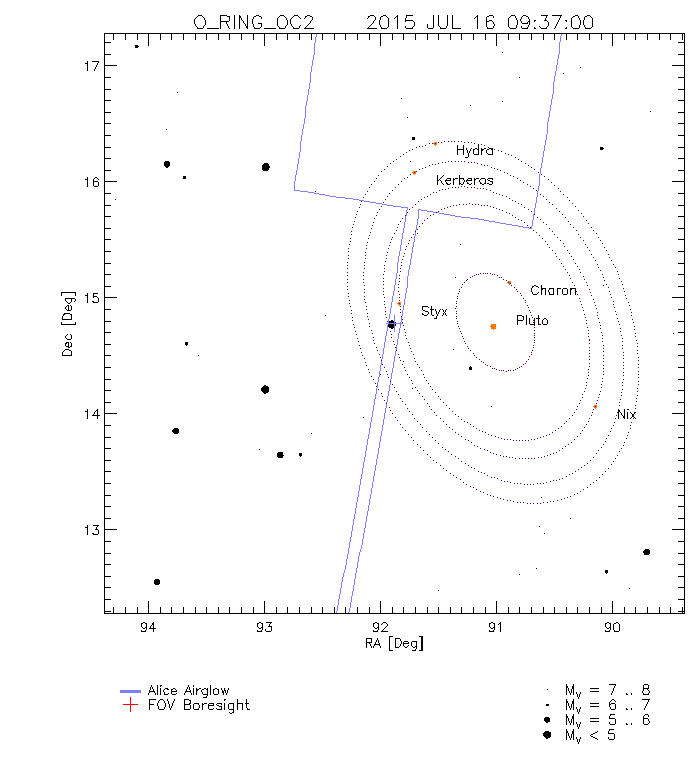}
\caption{Geometry of the $\texttt{O\_Ring\_Oc2}$ occultation of the region of Nix's orbit.}
\label{fig:geometry_alice_outer}
\end{figure}

The first set of occultations (\texttt{Alice\_StarOcc1},
Figure \ref{fig:geometry_alice_inner}) was observed starting approximately 4\,h
after the closest approach to Pluto.
Alice monitored two bright stars, which were placed in
the slit simultaneously, allowing two different occultations to be measured with
one sequence. The observation lasted 2.5~h, and Alice data was read out at
250~Hz, resulting in $2.25\times10^6$ independent spectral images.
The first star's
projected distance ranged from $0.5r_{\rm P}$ to $1.9r_{\rm P}$ (600--2260 km) from Pluto's center,
where $r_{\rm P}=1188.3\pm1.6$ km is Pluto's radius \citep{nimmo}.
(\textit{i.e.}, a full Pluto occultation), while the other covered
an appulse spanning $1.5r_{\rm P}$ to $2.5r_{\rm P}$ (1800--3000 km).
The projected velocity varied
due to the rapidly changing geometry, but was typically 0.5 km/s to 1 km/s,
giving us meter-scale sampling in many cases.  The Fresnel limit $r_F$ was
4-5~m. Additional details of the observing geometry are given in
Table~\ref{tab:alice_geometry}.

\begin{deluxetable}{lcccrrcl}
\tabletypesize{\scriptsize}
\tablecolumns{8}
\tablewidth{0pt}
\tablecaption{Alice stellar occultation geometry.
Distances are projected distance from star to Pluto center.}
\tablehead{\colhead{}&\colhead{}&\colhead{}&\colhead{\bf Projected}&\colhead{}&\colhead{}&\colhead{}&\colhead{}\\
\colhead{}&\colhead{}&\colhead{}&\colhead{\bf Velocity}&\colhead{\bf Inner}&\colhead{\bf Outer}&\colhead{}&\colhead{}\\
\colhead{\bf Sequence}&\colhead{\bf Start}&\colhead{\bf Inner}&\colhead{\bf [km/s]}&\colhead{\bf [km]}&\colhead{\bf [km]}&\colhead{Tilt}&\colhead{\bf Star}}
\startdata
\texttt{Alice\_StarOcc1} & 2015 Jul 14 15:53 - 18:23 & 9000.0 s & 0.79    & 322  & 6,111 &$45\adeg6$ & HD 43153 \\
"                  & "                    & "        & 0.66  & 2,411 & 6,858  &$45\adeg6$ & HD 42545 \\
\texttt{O\_Ring\_Oc2}       & 2015 Jul 16 09:37 - 10:37 & 3600.0 s  & 0.39  & 46,794 & 48,185 &$43\adeg4$ & 67 Ori   \\
\texttt{O\_Ring\_Oc3}       & 2015 Jul 16 21:25 - 22:24 & 3593.4 s  & 0.39  & 63,233 & 64,625 &$43\adeg4$ & 67 Ori \\
\enddata
\end{deluxetable}
\label{tab:alice_geometry}

The inner Pluto-Charon region probed by these occultations was particularly
interesting because it sampled the region
where dust grains can survive for at least $10^4$ Pluto-Charon rotation
periods (or 175 years) \citep{gwv13}. \citet{ksw16} have analyzed this data set
to probe Pluto's atmospheric structure; we use the data
far enough from Pluto that it is not affected by atmospheric refraction.

\begin{figure}[htbp]
\centering
\includegraphics[keepaspectratio,width=4 in]{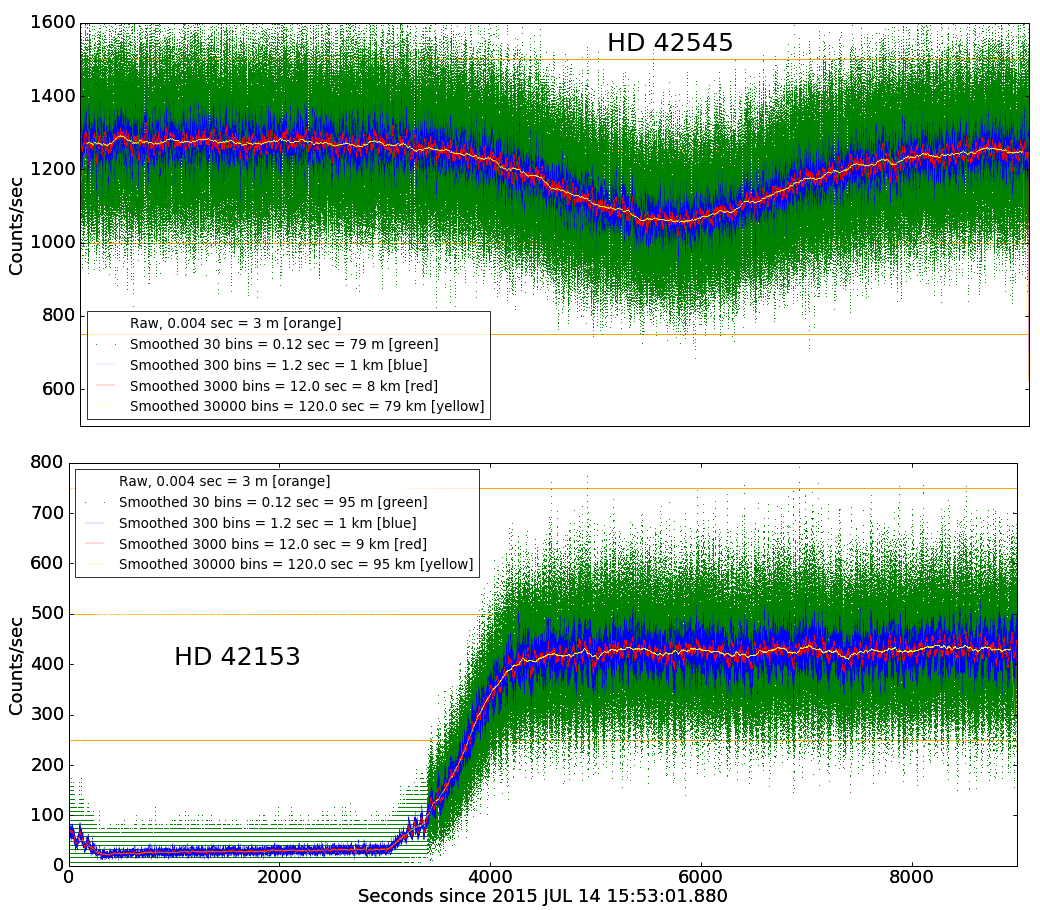}
\caption{Light curves from the Alice occultations of the Pluto-Charon region,
both taken in the \texttt{Alice\_StarOcc1} sequence.  The orange points
(which appear as discrete horizontal lines due to their low individual
DN values) are the raw observations at 250 Hz,
and the other curves are binned to larger spatial scales.
The broad dips are due to the appulse and occultation of Pluto.}

\label{fig:data_alice_inner} \end{figure}

\begin{figure}[htbp]
\centering
\includegraphics[keepaspectratio,width=4 in]{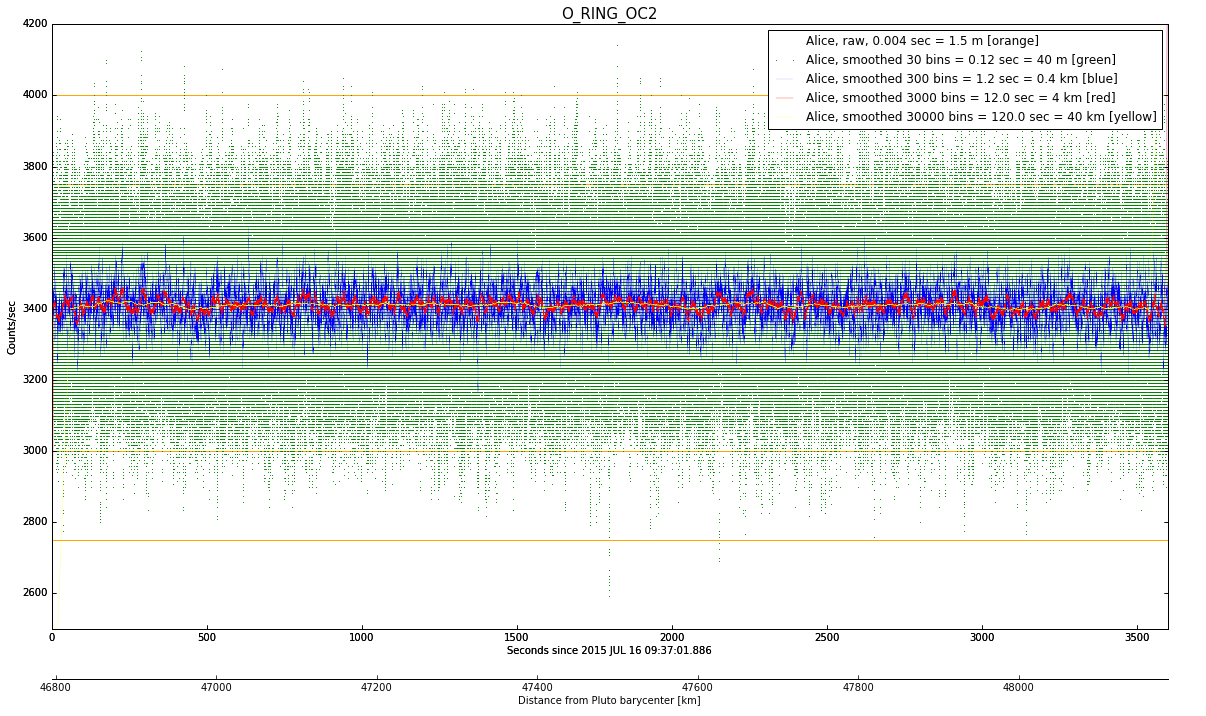}
\caption{Alice occultation in the Nix region, $\texttt{O\_Ring\_Oc2}$.
The curve for $\texttt{O\_Ring\_Oc3}$ is similar, but is not shown.}
\label{fig:data_alice_outer}
\end{figure}

The second and third occultations occurred approximately two days after the
Pluto encounter. These sequences --- \texttt{O\_Ring\_Oc2} and
\texttt{O\_Ring\_Oc3} --- probed the region around the orbits of Nix and Hydra,
respectively, where collisional debris from these bodies might be found
(Figure \ref{fig:geometry_alice_outer}).  Dynamical studies show that dust in
these regions can be stable for at least $10^5$ Pluto rotation periods
(1,750 years) \citep{gwv13}. Because these observations were taken later during the
encounter, their projected speeds are much lower ($\sim0.37$~km/s), and the
radial width covered by each scan is just slightly over $1r_{\rm P}$ (1188 km).  Typical
Fresnel resolution $r_F$ was $\sim10$~m, and per-sample resolution at 250~Hz was
1.5~m.

We analyzed all these Alice occultations in the same way. We started with the
raw `pixel list' data, which enumerates all photons received during the
occultation, time-tagged to 4ms resolution. We used this
information to extract a light curve, incorporating only the photons received on
the appropriate rows corresponding to the target star location.
We summed spatially over three rows. We then binned photons at all wavelengths,
omitting 20 adjacent columns which were contaminated by
interplanetary H Ly-$\alpha$ emission near 1216~\AA.
The remaining signal was the light curve $I(t)$.
The typical count rate was 400--3400 DN/s, depending on the target star.

Stellar flux during \texttt{Alice\_StarOcc1} was affected by vignetting as
the slit's sky position moved within its $0\adeg035\times0\adeg035$
box.  This motion by the spacecraft's guidance system was not enough to move the
star off the pixel, but did reduce the flux by about 50\% at the extremes of
motion. We corrected for this by dividing the light curve $I(t)$ by a simple
linear vignetting model based on the spacecraft's pointing in declination.
Vignetting in the RA direction was negligible.

The per-sample resolution of these observation is below the Fresnel limit.
We therefore binned the data using boxcar smoothing, to search for occultations
on a variety of scales from 10~m - 100~km. Binning allowed us to
search for individual bodies at meter scales,
in addition to extended clouds or rings of
material at larger spatial scales. Light curves from the Alice occultations are
shown in Figs.~\ref{fig:data_alice_inner}-\ref{fig:data_alice_outer}.

The optical thickness of a ring can be computed by comparing the flux at each
timestep $F$, to the mean flux $F_0$, 
\begin{equation}
F = F_0 \ \exp(-\tau / \mu),
\end{equation}
Usually we refer to the normal optical
thickness $\tau_n$, where $\tau_n = \tau \mu$.
Our limits on optical depth are given in Table~\ref{table:results_alice}.
We did not find any statistically significant features
in any of the four occultation curves.
At the fine-resolution limit, Alice would have detected individual bodies of
size 5-10 meters; it detected none along its path. When binned at larger
resolutions, the data would have easily detected a Chariklo-type ring
\citep{chariklo} in the Pluto system ($\tau = 0.4$, width = 5~km), but such a
feature was not seen.  Binned to low resolution,
Alice's limit of $\tau_n < 0.006$
at 40 km does not exclude the existence of a Jupiter-type ring at Pluto,
although such a ring was ruled out by the LORRI and MVIC imaging data described
elsewhere in this paper.

\begin{table}
\centering
\caption{Alice optical depth limits from stellar occultations.}
\begin{tabular}{lrl}
\textbf{Sequence}  & \textbf{Resolution} & \ \ \ $\tau_n$  \\\hline
\texttt{O\_Ring\_Oc2}; \texttt{O\_Ring\_Oc3}   & $r_F = 11$~m & $<$0.3 \\
                                               & 50 m       & $<$0.1  \\
                                               & 500 m      & $<$0.03 \\
                                               & 5 km       & $<$0.01 \\
                                               & 50 km      & $<$0.003 \\\hline
\texttt{Alice\_StarOcc1}                       & $r_F = 4$~m  & $<$0.43 \\
HD 42545                                       & 0.08 km    & $<$0.20  \\
                                               & 0.8 km     & $<$0.065 \\
                                               & 8 km       & $<$0.028 \\
                                               & 80 km      & $<$0.0080 \\\hline
\texttt{Alice\_StarOcc1}                       & $r_F = 4$~m  & $<$0.24 \\
HD 42153                                       & 0.1 km     & $<$0.14  \\
                                               & 1 km       & $<$0.049  \\
                                               & 10 km      & $<$0.024  \\
                                               & 100 km     & $<$0.0090 
\end{tabular}
\label{table:results_alice}
\end{table}

Together these occultations explored only about 15\% of the total radial extent
between Pluto and Hydra.  Our integration length was limited by the spacecraft's
downlink scheduling and Alice's relatively large data volume. The imaging data
covered this region and others completely, although at lower resolution than
Alice's finest limit.

\clearpage
\section{Direct Detection of Dust Impacts}

The Student Dust Counter (SDC) onboard NH
was designed to directly detect the impact of dust particles \citep{sdc}.
SDC measures the mass of dust grains in the range of $10^{-12}<m<10^{-9}$ g,
covering particle radii of approximately 0.5 to 10 $\mu {\rm m}.$
A period of $\pm5$ days centered on Pluto closest approach,
corresponding to approximately $\pm5000r_{\rm P}$ ($5.9\times10^6$~km)
was used to evaluate the dust density near Pluto.
Throughout this entire period, SDC recorded only a single event
that could be attributable to a dust impact \citep{bagenal}.

During the Pluto encounter, propulsive spacecraft
attitude maneuvers executed by New Horizons
introduced excessive noise to the SDC detectors.
To avoid recording this noise, SDC was set to a charge threshold of $Q=10^7e^-$
(where $e^-$ is the electron charge),
corresponding to a smallest detectable particle radius of $\sim1.5~\mu{\rm m}.$
From the amplitude distribution of all recorded noise events near
Pluto encounter, it was determined this event was above the $2\sigma$
error threshold of SDC noise events within $\pm5000r_{\rm P}.$
Hence, the probability that this event was due to a true dust particle
impact was estimated to be $\sim95\%.$

As described in \citet{bagenal}, this event was used to define an upper
limit for the dust density within $5000r_{\rm P}$ for grains larger
than 1.4 $\mu{\rm m};$ from this calculation we found the dust density
for particles larger than 1.4 $\mu{\rm m}$ to be $\sim1.2~{\rm km^{-3}}.$
A 90\% confidence level for the dust density in this region of
space was calculated to be in the range $0.6 <n<4.6~{\rm km^{-3}}.$
However, we caution that our single candidate particle impact
occurred at a distance of $\sim3070r_{\rm P}$ ($3.65\times10^6$~km).
This great distance, combined with (1) the fact that the
spacecraft was then very far above the equator plane and (2) that no other particle
impacts were detected closer or in this plane,
make it unlikely that this particle is associated with rings
or other coherent dust structures in the Pluto system.
Further, within the error bars the densities at the high threshold setting of
SDC (1.4 $\mu$m radius) before, during, and after the encounter remained
the same (2015-2017), indicating the detection near
Pluto was likely not special.
We conclude that this particle was most likely a background
interplanetary dust particle from the Kuiper Belt. 

\section{Searching by Forward-Scattered Light}

\subsection{Design of the Observational Program}

After closest approach, New Horizons looked back
at the Pluto-Charon system at several epochs to search for faint dust rings
through forward-scattered sunlight.  Both LORRI and the
Multi-spectral Visible Imaging Camera (MVIC), a component of the
Ralph remote sensing package,
were used to generate deep image-mosaics covering a
large angular area of the sky centered on Pluto.
The forward-scattering observation sequences for both LORRI and MVIC
are listed at the bottom of Table \ref{tab:obs}.
LORRI was described in $\S\ref{sec:back}$ in the context of looking
for rings in back-scattered sunlight.  In brief, MVIC comprises
a set of narrow strip-format CCDs with various band-passes.
For the ring search,
we used the MVIC panchromatic (Pan-Frame) CCD, which is a $5024\times128$
pixel array used in white-light (400 - 975 nm bandpass) at the
focus of a 7.5 cm off-axis reflector.
Unlike the other MVIC CCD channels, which scan over the field
of view using time-delay integration,
MVIC Pan-Frame uses a standard frame-transfer CCD.
The pixel scale is $4\asec1,$ yielding a $5\adeg7\times0\adeg15$ field.
See \citet{mvic} for further details.

\begin{figure}[htbp]
\centering
\includegraphics[keepaspectratio,width=4 in]{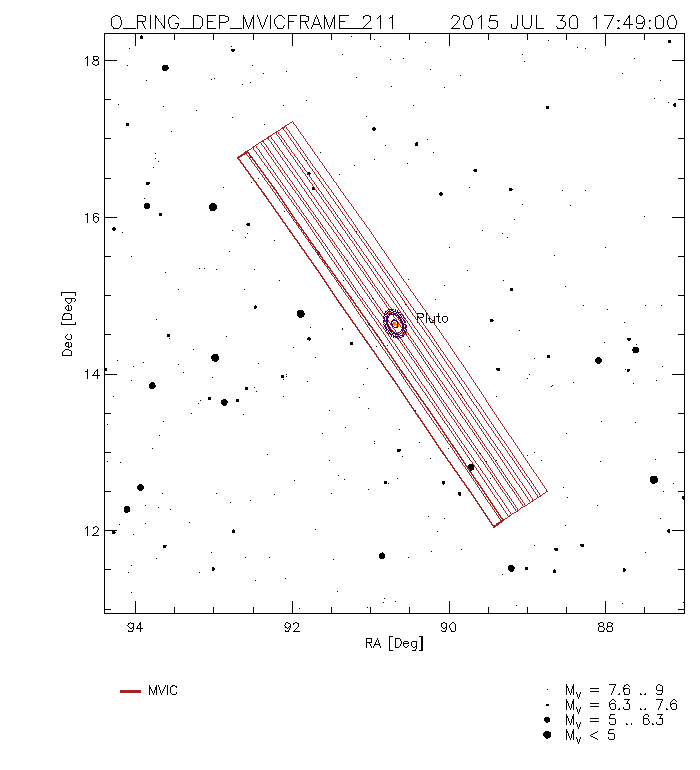}
\caption{The footprint on the sky for the
\texttt{O\_RING\_DEP\_MVICFRAME\_211} MVIC forward-scattered light
search on P$+16.$
Note that the mosaic is aligned with and elongated along the projected
major-axis of the orbits of the minor satellites.
The same footprint is also used for the P$+7$ and P$+110$ searches.}
\label{fig:mvic_foot}
\end{figure}

\begin{figure}[htbp]
\centering
\includegraphics[keepaspectratio,width=5 in]{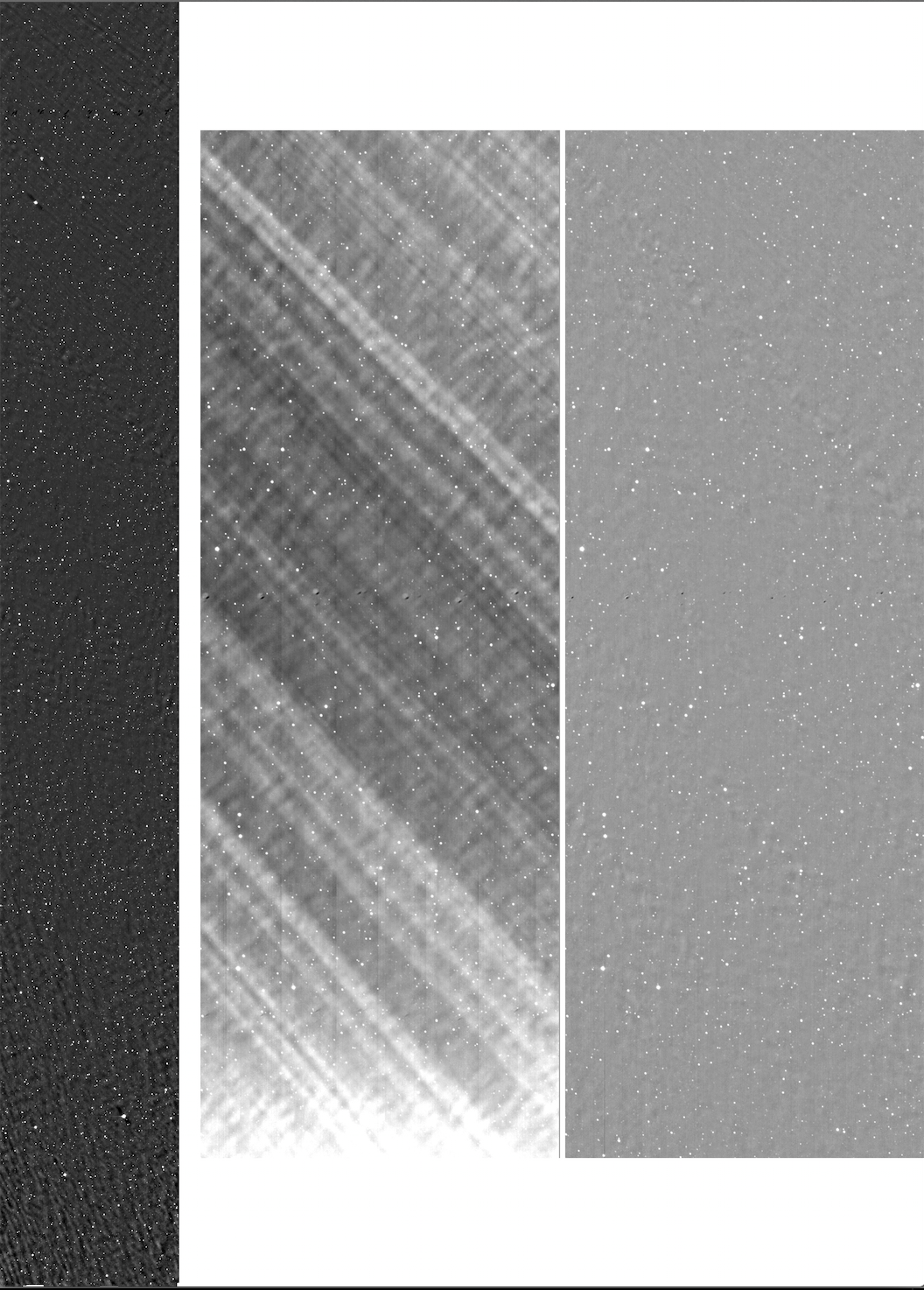}
\caption{Left: The complete reduced stack of the P$+110$
 MVIC outbound mosaics. The field covered is $5\adeg7\times0\adeg8$.
Middle:  The central portion of the first day-P$+110$ mosaic zoomed by
$2\times$ to show the strong stray sunlight pattern.
Right:  The same region corrected for stray light.}
\label{fig:mvic_red}
\end{figure}

\begin{figure}[htbp]
\centering
\includegraphics[keepaspectratio,width=6 in]{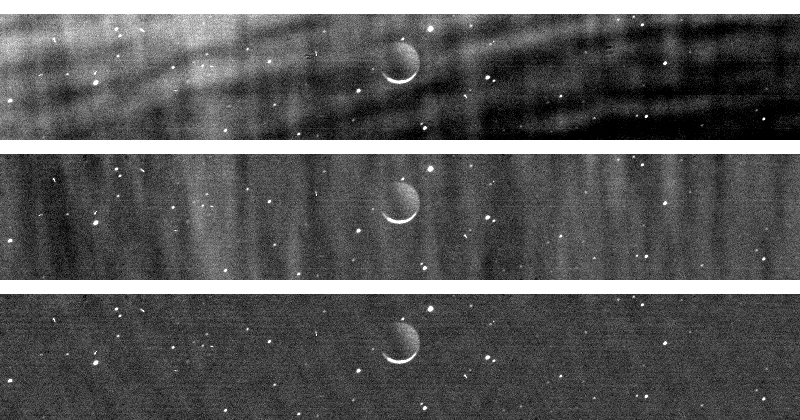}
\caption{Image reduction demonstrated for a single image taken from
the MVIC outbound mosaic \texttt{O\_RingDep\_A\_1}. Top: The central portion of
the image within the mosaic that included the ``nightside''
hemisphere of Charon. The strong mottling
pattern is due to stray sunlight entering the camera.
Middle: The first reduction step is to remove
the average pattern seen in at all images within the mosaic.
Bottom: The second reduction step is to remove the streaks that
traverse the entire mosaic.  Note the recovery of
``Pluto shine'' on the upper right portion of the dark hemisphere of Charon.}
\label{fig:mvic_reduc}
\end{figure}

In general design, the LORRI and MVIC searches have strong similarities.
Both instruments were used within an hour of each other at given
epoch and covered roughly similar areas.  Further, as was done for
the hazard searches, LORRI was again used with $4\times4$ binning
to maximize sensitivity and minimize telemetry; as such the LORRI
and MVIC pixel scales were nearly the same.  Apart from
an initial search conducted using MVIC only a day after
closest approach, joint LORRI and MVIC
searches were conducted at days P$+7,$ P$+16,$ and P$+110.$
The final epoch extended the search to the full Hill sphere surrounding Pluto,
where $r_{\rm Hill} = 6.4\times10^6~{\rm km}$ at Pluto
($99.3\times$ the orbital radius of Hydra, for comparison).
The phase angle was nearly constant at $\sim165^\circ$
over all epochs, and the close angular proximity of the sun
introduced strong stray light into both cameras.
The MVIC searches, however, were nearly two orders of magnitude
more sensitive than those done with LORRI
due to the much smaller effects of stray light in MVIC, plus the
longer integrations done with this instrument.
We detail the LORRI and MVIC sequence designs further
in the next two subsections.

\subsection{The MVIC Mosaics}

\subsubsection{Observation and Preparation of the MVIC Mosaics}
\label{sec:mvic}

\begin{figure}[h]
\centering
\includegraphics[keepaspectratio,width=5 in]{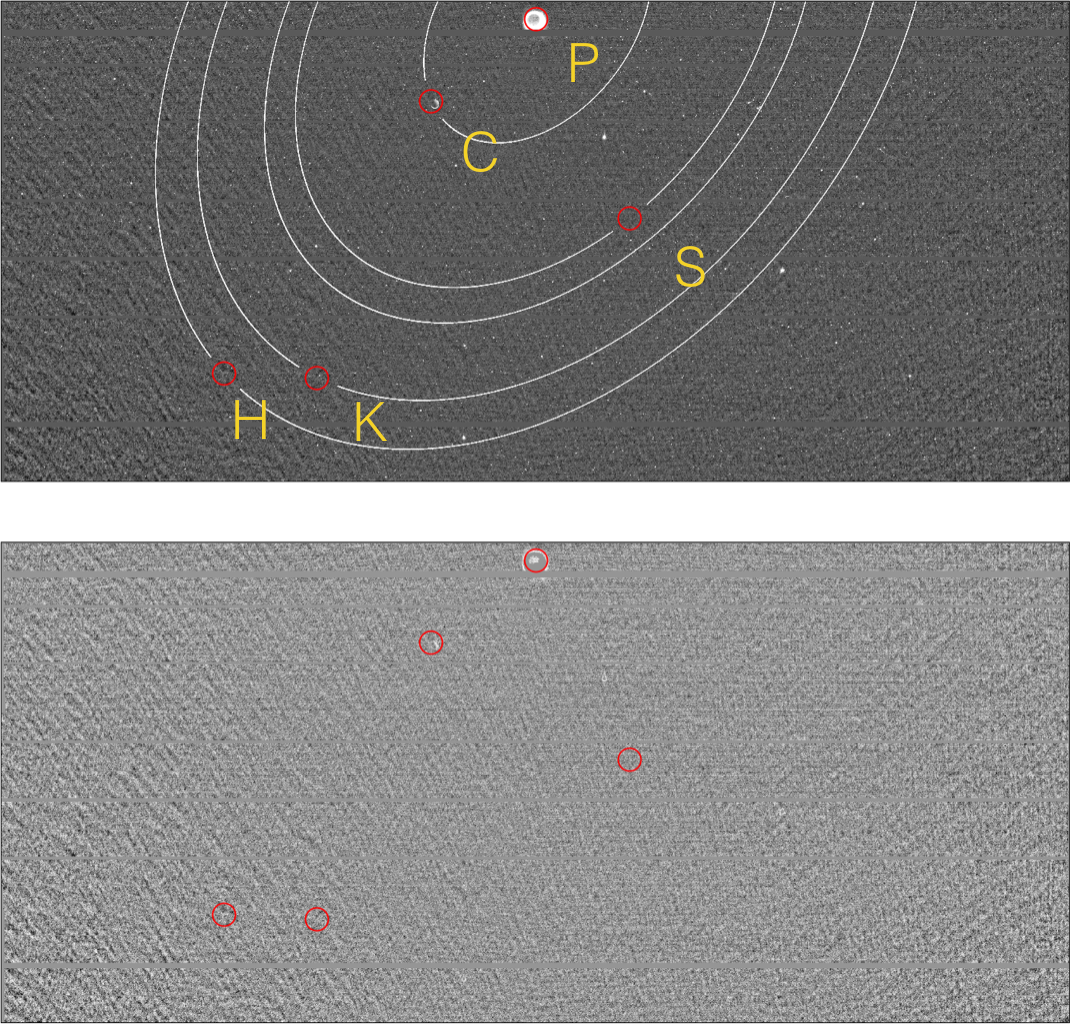}
\caption{MVIC outbound mosaic \texttt{O\_RingDep\_A\_1}, taken at P+4~h.
Top: MVIC mosaic, stray-light-subtracted,
with orbits of Pluto system indicated (P=Pluto, and so forth).
Bottom: Same as top, but with rescaled with bright pixels removed
so as to reveal any structure in the background image.}

\label{fig:mvic_data}
\end{figure}

The MVIC forward-scattering ring search sequences benefit
from markedly longer exposure times that go far beyond compensating
for the smaller aperture of Ralph as compared to LORRI, a markedly
lower amplitude of stray sunlight, and fortuitously,
a pattern of stray light that is considerably easier to remove.
The approach to constructing the MVIC mosaics was to align the long axis
of the CCD with the projected major axis of the satellite orbits,
using spacecraft pointing to tile the field in the minor-axis direction
(Figure \ref{fig:mvic_foot}).  For the P$+7,$ P$+11,$ and P$+110$ mosaics,
images were obtained at 7 pointings along the minor-axis of the projected
orbital plane, with $25\times10$s exposures obtained at each location,
for a 250s total exposure.  The reduced P$+110$ mosaic is shown
in Figure \ref{fig:mvic_red}, as well as a central
portion of it before and after the stray sunlight was corrected for.
The exposures were split into separate sequences
of 20 and 5 exposures, with a small offset of a few pixels between the two.
The images at any position also overlapped slightly with those at
flanking positions.\footnote{In passing we also note that we actually
obtained 21 and 6 exposures at each location, but the first exposure in MVIC
Pan-Frame sequences always suffers from large motion blur and is discarded.}
For the P$+110$ search, as with LORRI, the full mosaic was done three times
for 750s total exposure.  The first MVIC search, obtained
a day after close approach, comprises 17 pointings along the projected
minor axis, with $15\times10$s exposures or 150s total at each.  This mosaic
is offset, with Pluto positioned near the edge in the minor axis direction,
and the spacing of some of the pointings is such that the images
do not overlap, leaving gaps in portions of the mosaic.

Stray sunlight affected the MVIC images, but at a lower level and with less
highly variable structure than was seen in the LORRI images.
Examination of the complete ensemble of MVIC images for a given mosaic
showed that the stray light pattern could be decomposed into a
constant component that was present in all images,
and a pattern of streak or rays that appeared to be fixed in space.
The constant component could be visualized by
averaging the images, although in practice we isolated it by a PCA
analysis of the ensemble, iteratively rejecting stars
from affecting the extraction of the component (cosmic rays were
easily identified and repaired from the multiple images
available at each position).
Once derived, the constant component could be subtracted from all images
in the ensemble, leaving the pure streak-pattern behind.
Looking at a complete mosaic, it was clear that the streaks appeared to be
radiating from a point on the sky (although this was not the
projected location of the sun).
The streaks were then removed with an {\it ad hoc} algorithm that smoothed
along radial lines emanating from the streak radiant.  The smoothing
length was $\sim80$ pixels and the operation was conducted on the
mosaiced images with stars clipped out.  These reduction steps are demonstrated 
Figure \ref{fig:mvic_reduc}.

\subsubsection{Ring limits from the MVIC Observations}

An example of the processed MVIC ring mosaics is shown in
Figure~\ref{fig:mvic_data}.
Here ${\rm R_{\rm H}}$ refers to the distance of Hydra
from the system barycenter, 64,738~km.
No rings are obvious in the mosaics.
But planetary ring structure can often exist
at levels that are not easily visible to the eye, and in many cases they can be
seen when many pixels are summed appropriately \citep[\textit{e.g.,}][]{sc93, te98}.
We used this technique to search quantitatively for faint rings that might be
unseen to the eye.

\begin{figure}[htbp]
\centering
\includegraphics[keepaspectratio,width=4 in]{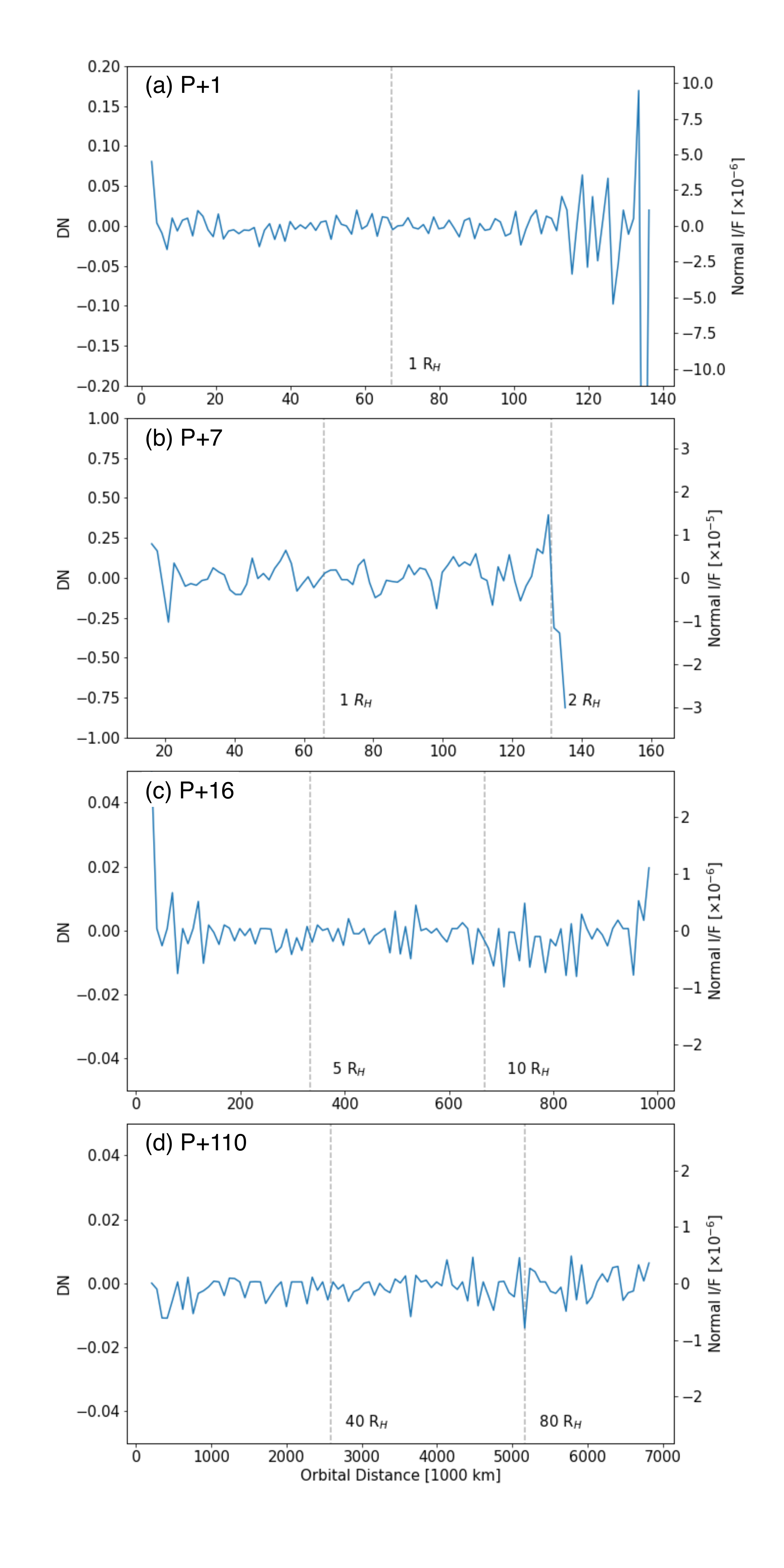}
\caption{Radial profiles from the MVIC mosaic at (a) P$+1$, (b)  P$+7$, (c) P$+16$, and (d) P$+110$.
Pixels are binned by radial distance from the system barycenter,
using N=100 radial bins. No ring structure is apparent.
The larger scatter at the extreme inner and
outer regions is due to larger errors in the stray-light correction at the
margins of the detector and is not significant.}
\label{fig:mvic_profile}
\end{figure}

We first registered the full mosaics precisely based
on the known positions of field stars. This corrected for the
several-pixel uncertainty in our NH pointing knowledge.
Based on the measured pointing, we then generated backplanes for the images,
where for each pixel we computed a value for the projected
radial distance and azimuth angle, referenced relative to a plane
centered on Pluto and normal to Pluto's pole \citep{saa07}.
We performed a two-pass removal of any bright pixels with flux exceeding
$3\sigma$ above the median, which were typically due to stars or known bodies
in the Pluto system. We then grouped pixels by radial distance, and computed the
mean within each bin. We used both N=100 and N=1000 radial bins, spaced linearly
from zero to the maximum radial distance in each mosaic.
Radial profile $I/F$ profiles from all four epochs of MVIC observations
are shown in Figures \ref{fig:mvic_profile}a to \ref{fig:mvic_profile}d. 

The radial profiles did not reveal any features that were clearly indicative of
rings or arcs. There were no features that appeared aligned with the projected
geometry of the Pluto system.
All visible structures in the images appeared to be
associated with either stray light, or the edges of our mosaic scan pattern.

Our radial profiles can be used to place quantitative upper limits on ring
material at Pluto. We first converted from DN to $I/F$ using the
instrumental calibration constant \texttt{RSOLAR}:
\begin{equation}
I/F = C\, /\,  \texttt{RSOLAR},
\end{equation}
where
$\texttt{RSOLAR} = 9.8813\times10^{4}{\rm DN/s}~({\rm erg~cm^{-2}~s^{-1}~\AA^{-1}~sr^{-1}})^{-1}$
for MVIC's Pan-Frame sensor, and $C$ is the count rate in DN/s.
As before, the solar flux at the MVIC Pan Frame pivot wavelength
($\lambda$ = 6920~\AA) was taken to be $F=145{\rm~erg~cm^{-2}~s^{-1}~\AA^{-1}}$
at 1~AU.  Our upper limits for normal $I/F$ are
given in Table~\ref{table:results_lorri_mvic}. The most sensitive limits are for
the $\texttt{O\_RING\_DEP\_MVICFRAME\_305}$ observations
(Figure \ref{fig:mvic_profile}d), for which we found a
normal $I/F$ limit of $< 7.3\times10^{-7}$ when binned to 100 radial bins at a
resolution of 68,140 km. At 750s/pixel, this observation had the longest
integration time of all of our outbound imaging.  The other MVIC mosaics had
shorter exposure times, and correspondingly higher $I/F$ limits.

\subsection{The LORRI Mosaics}

\subsubsection{Observation and Preparation of the LORRI Mosaics}

The general strategy at all three LORRI search-epochs was to map out a 15-point
mosaic using three strips of overlapping LORRI field.  The mosaic geometry
was roughly patterned to align with the highly elliptical projected-orbits
of the known satellites on the presumption that the rings were also likely
to have circular orbits within the common orbital plane of the Charon and
the minor satellites.  The central
strip of the mosaic comprised seven LORRI fields and covered the projected
major axis of the satellite orbits.  This was flanked by parallel strips of four
fields on both sides, producing a symmetric pattern.  Given the even
number of fields in the flanking strip versus the odd number in the central
strip, the boundaries between fields in the flanking strips and those
in the central strip were offset by half of the LORRI field.

\begin{figure}[htbp]
\centering
\includegraphics[keepaspectratio,width=5 in]{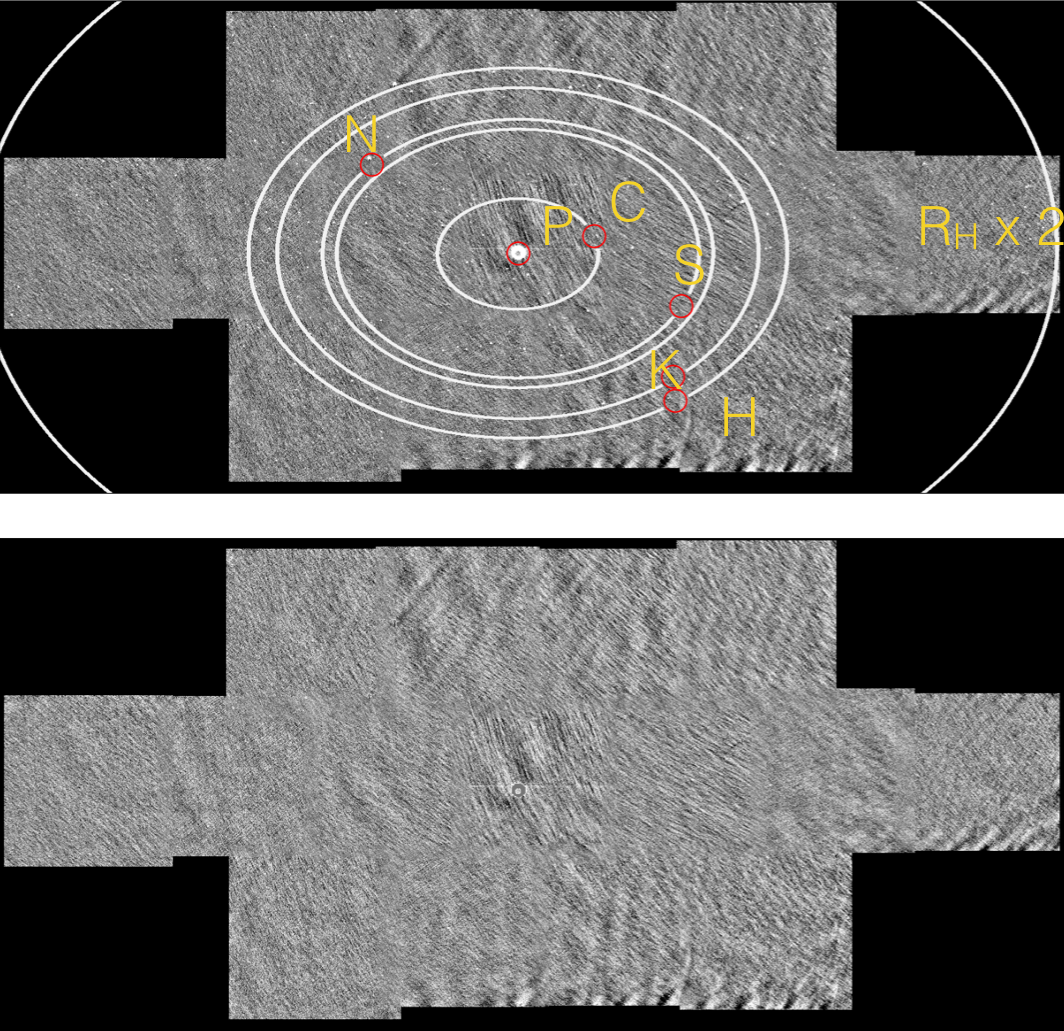}
\caption{Top: LORRI mosaic $\texttt{O\_RING\_DEP\_LORRI\_202}$, overlain with
orbits of Pluto's satellites, with their positions marked (P=Pluto, and so
forth). Bottom: Same, but with stars and Pluto-system bodies masked. The
remaining structure is due to incompletely removed solar stray
light.}
\label{fig:lorri_data}
\end{figure}

For the day P$+7$ and P$+16$ sequences,
$20\times 0.4$s exposures were obtained at each of the 15 positions, giving
an 8s total exposure, although this is effectively doubled in
the significant overlap areas (roughly $\sim15\%$ of the
field in both dimensions) between adjacent fields.  On day P$+110$
the mosaic pattern was repeated 3 times,
although with $15\times0.4$s and $5\times0.2$s
exposures at each pointing for a total integration of
$3\times(15\times0.4+5\times0.2)=21$s at any position (ignoring
the overlap of adjacent fields).

\begin{figure}[htbp]
\centering
\includegraphics[keepaspectratio,width=5 in]{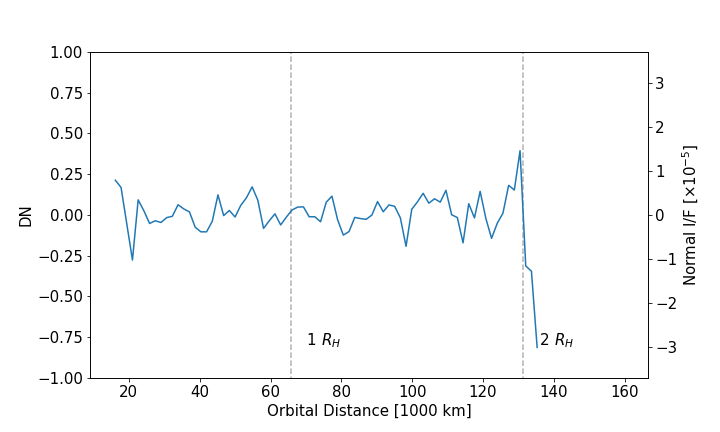}
\caption{Radial profile from the \texttt{O\_RING\_DEP\_LORRI\_202} mosaic.}
\label{fig:lorri_profile}
\end{figure}

As noted above, the LORRI mosaics were heavily compromised by the
strong stray sunlight.  The primary limitation is that the
absolute intensity level of the stray light was a significant
fraction of the CCD full well, requiring the exposures to be limited
to only 0.4s, in contrast to the markedly deeper 10s exposures conducted
during the hazard searches.  Subtraction of the stray light
was also problematic.  The pattern of background light in any
image was highly structured with features that changed rapidly
with even small changes in spacecraft pointing with respect to the sun.
We hoped that a PCA approach applied to the entire ensemble
of LORRI search images could allow an accurate stray light model
to be constructed for any given image, but unfortunately the behavior
of the stray light had a component that was unique to any position,
such that it could only be partially modeled by other images in the ensemble.
PCA models did remove a portion of the stray light,
but could not fully correct for it.  Later analysis of the MVIC
mosaics (see $\S\ref{sec:mvic}$)
motivated a general-purpose algorithm for removing the
ray-like streaks seen in some of the structure of the stray light,
which did improve the reduction of the LORRI mosaics, but even perfect
correction of the stray light would still leave behind the
strong shot-noise associated with it.

We emphasize a strong caveat that was always kept in mind during the analysis
of the forward-scattering mosaics was to not perform any reduction steps that
would also eliminate real diffuse-features associated with dust rings or arcs.
If, for example, one was only interested in compact point sources
in the fields (although looking back at the system we presumed
that any unknown minor satellites would be invisible), a high-pass spatial
filter would quickly eliminate most of the stray light structure,
but it would do so
at the expense of any real well-resolved dust features also present.
One could also attempt PCA on only the 20 (or 60 images for the day P$+110$
searches) at any position, but here the spacecraft pointing variations
were relatively small, which again would mean that there would be
little discrimination between real ring-features from
the structure of the stray light.

\subsubsection{Ring limits from the LORRI Observations}

Despite these difficulties, we were able to produce reduced LORRI
mosaics at each epoch that could be searched for forward-scattered
light from any dust rings.  We analyzed the outbound LORRI
observations in much the same way as the MVIC
observations. The mosaiced images were registered using field stars; we then
created geometrical backplanes referenced to Pluto's orbital plane, removed
bright pixels, created radial profiles, and converted from DN to normal $I/F$.
We used the calibration constant
$\texttt{RSOLAR}=3.801\times10^{5} {\rm DN/s}~({\rm erg~cm^{-2}~s^{-1}~\AA^{-1}~sr^{-1}})^{-1}$
for LORRI 4$\times$4 frames.
An example image is shown in Figure \ref{fig:lorri_data},
and the resulting radial profile in Figure \ref{fig:lorri_profile}.

\begin{table}
\centering
\caption{Measured MVIC and LORRI normal I/F limits.}
\begin{tabular}{lrlrl}
\textbf{        }        & \multicolumn{2}{c}{\textbf{N=1000 radial bins}}  &
\multicolumn{2}{c}{\textbf{N=100 radial bins}}                 \\
        & \textbf{Resolution} & &
\textbf{Resolution} & \\
\textbf{Sequence}        & \textbf{(km)} & \textbf{Normal I/F} &
\textbf{(km)} & \textbf{Normal I/F} \\
\hline
\texttt{O\_RingDep\_A\_1} (MVIC)      &  136  & $<5.0<\times10^{-6}$ & 1,360   & $<1.9\times10^{-6}$ \\
\texttt{O\_RING\_DEP\_MVICFRAME\_202} &  432  & $<2.1\times10^{-6}$ & 4,320    & $<1.1\times10^{-6}$ \\
\texttt{O\_RING\_DEP\_LORRI\_202}     &  159  & $<2.8\times10^{-5}$ & 1,590    & $<1.4\times10^{-5}$ \\
\texttt{O\_RING\_DEP\_MVICFRAME\_211} &  983  & $<1.8\times10^{-6}$ & 9,830    & $<8.9\times10^{-7}$ \\
\texttt{O\_RING\_DEP\_LORRI\_211}     &       &  (not analyzed)  \\
\texttt{O\_RING\_DEP\_MVICFRAME\_305} &  6,814 & $<1.5\times10^{-6}$ & 68,140  & $<7.3\times10^{-7}$ \\
\texttt{O\_RING\_DEP\_LORRI\_305}     &  2,834 & $<1.3\times10^{-4}$ & 28,340  & $<1.2\times10^{-4}$
\end{tabular}
\label{table:results_lorri_mvic}
\end{table}

We did not find any features suggestive of rings in the LORRI images or radial
profile. Our lower limits are given in Table~\ref{table:results_lorri_mvic}.
The limits for LORRI are substantially worse (10-100$\times$)
than those for MVIC, for two reasons.
First, the LORRI exposures were much shorter, with the
integrated time for longest observations 18s/pixel, vs.\ 750s
for the MVIC observations at the same geometry.
The stronger LORRI background, with the poorer corrections noted
in the previous section, contribute the remaining lost of sensitivity.
Because the LORRI mosaics were taken on the same dates as MVIC,
but with shorter exposures and smaller fields,
we did not create the mosaic and radial profiles for the
remaining LORRI sequence \texttt{O\_RING\_DEP\_LORRI\_211}. 
The LORRI and MVIC observations were designed to be redundant,
and the remaining LORRI data would not improve our $I/F$ limits.

\begin{figure}[htbp]
\centering
\includegraphics[keepaspectratio,width=5 in]{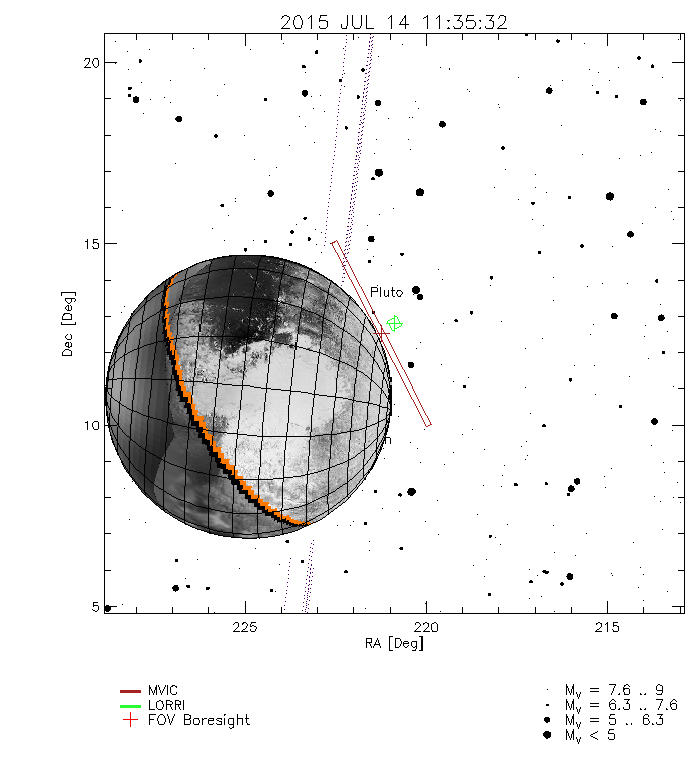}
\caption{The geometry of the $\texttt{PEMV\_01\_P\_MVIC\_LORRI\_CA}$
MVIC scan is shown. The scan occured shortly before the 
spacecraft crossed Pluto's equatorial plane. The field of MVIC is
the narrow rectangle position off the limb of Pluto.  During an MVIC
scan the spacecraft is slowly rotated, and the field is scanned
over the planet as the camera is read out using a time-delay integration mode.
Dashed lines mark the orbits of the minor satellites, which traverse
a small portion of the scan at only $\sim3^\circ$ inclination.}
\label{fig:mvic_mpan1}
\end{figure}

\section{Untargeted Ring Observations}

The previous sections described the explicitly
targeted ring-search observations.
It is possible that rings or debris may be present
and serendipitously detectable in other observations.
In particular, NH passed through Pluto's equatorial plane at
approximately 2015, July 14 11:37:10 UT (spacecraft frame).
Any ring material observed near this time
would appear brightened substantially due to the high line-of-sight optical
depth of a ring seen edge-on. 

The equatorial crossing occurred approximately 11 minutes before the Pluto
close-approach, so the cameras were observing Pluto's
surface at high resolution during this time.
Due in part to uncertainty in Pluto's position and the spacecraft's pointing,
some of the images recorded data off Pluto's limb. 
None of the LORRI image fields passed directly through the (nearby) lit
equatorial plane; however, the MVIC scan $\texttt{PEMV\_01\_P\_MVIC\_LORRI\_CA}$
obtained close to this time nominally includes a portion
of the plane at very low inclination in the background off the limb of Pluto.
The integration is short and the background is dominated by
scattering from haze particles in Pluto's atmosphere;
no features associated with the orbits of the minor satellites were seen
(Figure \ref{fig:mvic_mpan1}.
We individually inspected all images around this time to search for any evidence
of ring material, and didn't find any. Because the line-of-sight did not
directly intersect the expected ring plane, any material we did find would have
been inclined. Since we cannot constrain the inclination or radial
extent of these non-detections, we cannot use these to measure a meaningful
upper limit.

\section{Discussion and Summary}

As Tables \ref{tab:hstlist} and \ref{tab:obs} show, substantial observational resources were
dedicated to the task of searching for dust, debris, or rings within the
Pluto-Charon system, using diverse tactics and instruments.
The NH science team in turn devoted substantial effort
to the reduction and analysis of these observations. Although
no rings or dust clouds were found, the program did produce significantly
improved upper limits on their allowed surface densities in the region of the
small moons.

\begin{figure}[htbp]
\centering
\includegraphics[keepaspectratio,width=5 in]{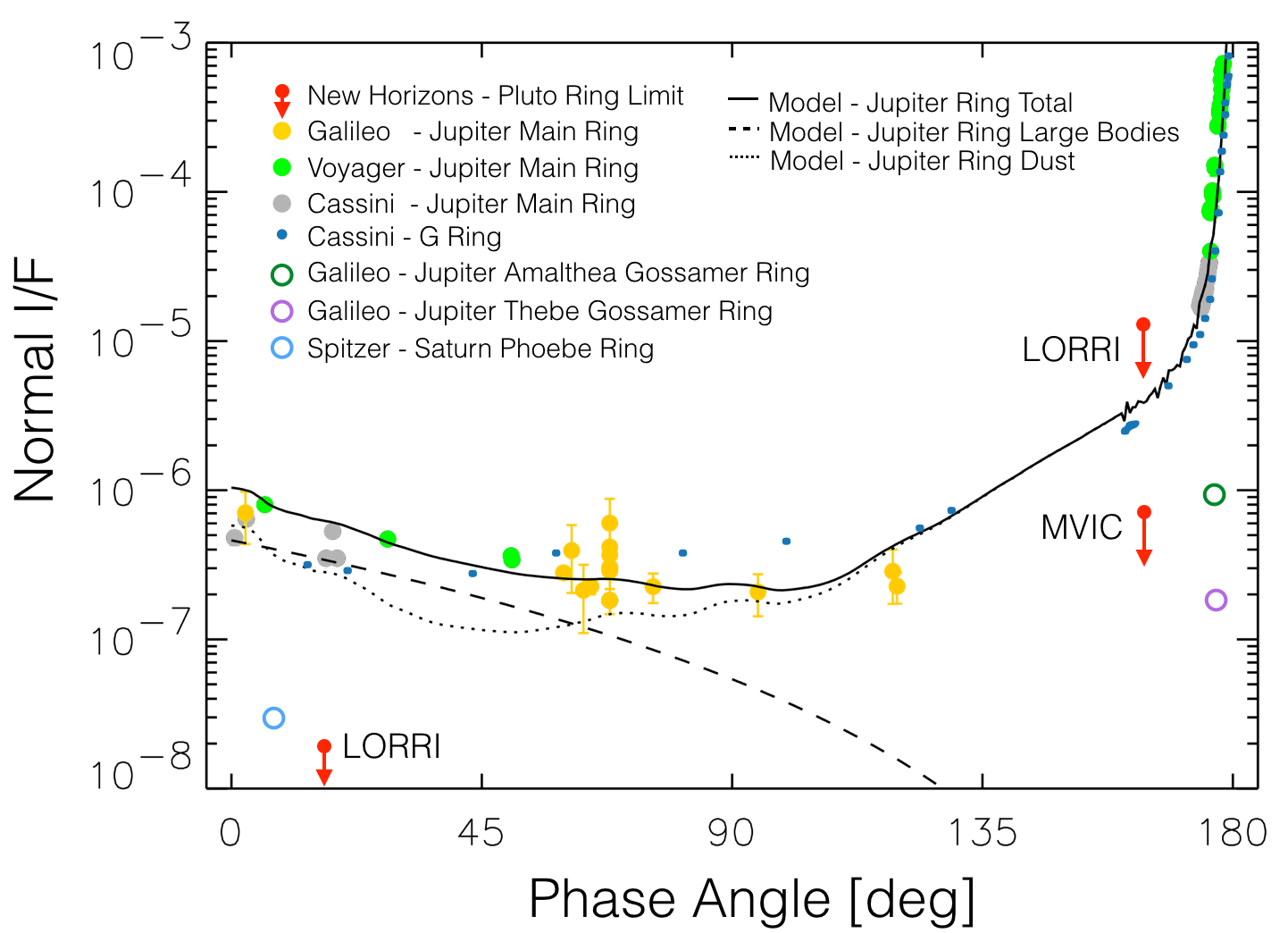}
\caption{Comparison between New Horizons detection limits and brightness of
several other rings. The $I/F$ (vertical axis) indicates relative reflectivity
at each ring's correct location. If Jupiter's main ring was placed in orbit
around Pluto, it would have been detectable on approach by LORRI,
and on departure by MVIC, despite the much lower flux at Pluto.
Saturn's G ring would also have been detected by LORRI and MVIC.
It is possible that NH could have barely detected the Phoebe
and Amalthea rings, if they were in orbit around Pluto.
On the other hand, Jupiter's Thebe ring is too faint to be detected.
Figure adapted from \citet{tpw04}.
G ring data from \citet{hedman}.
Voyager, Cassini, and Galileo data, and model ring, from \citet{tpw04}
and references therein.
Gossamer ring data from \citet{ockert};
Phoebe ring from \citet{verb}.}
\label{fig:nh_vs_jupiter_ring}
\end{figure}

In broad terms, NH
provided superior resolution and sensitivity to search for back-scattered
light from rings within the radial zone extending from the orbit of Styx,
the innermost satellite, to twice the orbital radius of Hydra,
the outermost satellite.  The new HST observations provided the best
constraints on broad features at large radii outside the satellite orbits.

At low phase angles, limits on the normal $I/F$ of narrow rings within this
zone are $2\times10^{-8}$ at the 1500 km resolution scale
of the P$-64$ images, decreasing to $7\times10^{-9}$ on $12,000$ km scales.
We can convert these limits to optical depth $\tau$ via Eq.~1, assuming that
the factor $A\times P(\alpha)/P(0)$ for rings is comparable to that for the
small moons: $0.7\pm0.2$ \citep{weaver}. If this assumption is correct, then
optical depth limits are $\sim30$\% larger than the
normal $I/F$ limits.

New Horizons also provided the first high-phase vantage
point to search for forward-scattered light from dust clouds or rings
during its departure from Pluto over the entire Hill sphere.
Here we derive limits of $\mu I/F<8.9\times10^{-7}$
on $\sim10^4$ km scales.

To aid in the interpretation of the these limits,
we demonstrate in Figure \ref{fig:nh_vs_jupiter_ring}
that we would have readily detected several of the
diffuse rings around Jupiter or Saturn,
had they been present at Pluto.
Jupiter's main ring would have been clearly seen, for example,
despite the 40$\times$ lower solar flux at Pluto vs.\ Jupiter.
Jupiter's main ring has a normal $I/F \sim 5\times10^{-6}$ and width
$\sim3000$~km \citep[\textit{e.g.,}][]{tpw04}.
Significantly, it comprises a mix of rocky debris and fine dust,
each component contributing a similar total cross-sectional area.
Our MVIC mosaics would have readily detected the dust in the
forward-scattering searches, at over an order of magnitude higher $I/F$
than our present limits, while the LORRI back-scattering searches would
have detected the debris component, with again an order of magnitude
sensitivity to spare.  At the same time, the Gossamer ring associated
with Jupiter's satellite Thebe would have remained undetected.
We cannot rule out the possibility that Pluto might
still have rings similar to the most tenuous ones known to exist.

As outlined in the introduction, prior to the New Horizons,
our expectations of whether Pluto could or even should,
have rings remained unsettled.
The discovery of four small satellites provided
a clear source for dust debris generated by gardening.
On the other hand, improved understanding of the role of solar radiation
pressure in clearing such material from the system \citep{solar},
coupled with the complex and richly packed orbital dynamics of the satellite
system \citep{youdin, sh15} mitigate against the existence of long-lived rings.

Recently, \citet{singer} found a
marked paucity of small craters on Pluto and Charon in NH images.
This would require
a dearth of KBO impactors smaller than $\sim1$ km in radius as compared to
the pre-encounter assumed impactor-size distribution.
In short, the smallest impactors appear to follow a markedly
shallower distribution function, such that their integrated
contribution to impact gardening may be decreased by well
over an order of magnitude below that
estimated by \citet{durda}.
This dearth of small-impactors is an additional
stroke against detectable rings.
The observational limits of the present work on the existence
of dust or rings in the Pluto-Charon system offer an
opportunity to revisit the theoretical expectations
on the on-going production and removal of dust within the Kuiper Belt.

\acknowledgments

We thank A. Steffl for help with the Alice data.
New Horizons observational geometry figures
were generated with GV \citep{throop_gv}.
This work has made use of \texttt{astrometry.net} \citep{lang}
and \texttt{astropy} \citep{astropy}.
This work was funded by the NASA New Horizons Project.
Additional support for MRS and DPH through Program number HST-GO-12436 was
provided by NASA through a grant from the Space Telescope Science Institute,
which is operated by the Association of Universities for
Research in Astronomy, Incorporated, under NASA contract NAS5-26555.

{}

\listofchanges

\end{document}